\documentclass[11pt,a4paper]{scrartcl}

\usepackage[english]{babel}

\usepackage[
  a4paper,
  left=25mm,
  right=20mm,
  top=20mm,
  bottom=30mm
]{geometry}

\usepackage{authblk}
\usepackage{amssymb}
\usepackage{booktabs}
\usepackage{amsmath}
\usepackage{color}
\usepackage{xcolor}
\usepackage{xspace}
\usepackage{diagbox,colortbl}
\usepackage{listings}
\usepackage{multirow}
\usepackage{graphicx}
\usepackage{eso-pic}
\usepackage{hyperref}
\usepackage{natbib}
\usepackage{caption}
\usepackage{subcaption}
\usepackage{makecell}
\usepackage{siunitx}
\usepackage{hyphenat}
\usepackage{tikz}
\usepackage{url}
\usepackage{changepage}

\newcommand*\circled[1]{\tikz[baseline=(char.base)]{
            \node[shape=circle,draw,inner sep=1pt,fill=yellow!40!white] (char) {#1};}}
\newcommand*\circledcyan[1]{\tikz[baseline=(char.base)]{
            \node[shape=circle,draw,inner sep=1pt,fill={rgb,255:red,128; green,255; blue,255}] (char) {#1};}}

\DeclareSIUnit\tdp{TDP}
\DeclareSIUnit\flop{FLOP}
\DeclareSIUnit\transfer{T}
\DeclareSIUnit\ethernet{E}
\DeclareSIUnit\channel{channel}
\DeclareSIUnit\channels{channels}
\DeclareSIUnit\iops{IOPS}

\newcommand{\colorCPU}{9EEBA5}
\newcommand{\colorGPU}{EB9F9F}
\newcommand{\colorFPGA}{9FD5EB}
\newcommand{\colorNetwork}{FFDA76}

\definecolor{listComment}{HTML}{008000}
\definecolor{listKeyword}{HTML}{2331FF}

\definecolor{tableblue}{HTML}{BBEBFD}
\definecolor{tableyellow}{HTML}{FFF5C3}
\definecolor{tablegreen}{HTML}{D7EEC9}
\definecolor{tablegray}{HTML}{EDEDFF}
\definecolor{tableorange}{HTML}{FFCB8D}
\definecolor{tablecpu}{HTML}{\colorCPU}
\definecolor{tablegpu}{HTML}{\colorGPU}
\definecolor{tablefpga}{HTML}{\colorFPGA}
\definecolor{tablenetwork}{HTML}{\colorNetwork}
\definecolor{n2papercolor}{HTML}{006400}

\newcommand{\myrowcolour}{\rowcolor[gray]{0.9}}%

\newcommand{\myrowcolourB}{\rowcolor[gray]{0.95}}

\newcommand{\myrowcolourCPU}{\rowcolor[HTML]{\colorCPU}}
\newcommand{\myrowcolourGPU}{\rowcolor[HTML]{\colorGPU}}
\newcommand{\myrowcolourFPGA}{\rowcolor[HTML]{\colorFPGA}}

\newcommand{\uproman}[1]{\uppercase\expandafter{#1}}

\newcommand{\OTUS}[0]{\textsc{Otus}\xspace}
\newcommand{\NOCTUATWO}[0]{\textsc{Noctua~2}\xspace}

\newcommand{\PCTWO}[0]{\textsc{PC2}\xspace}

\addto\extrasenglish{%
}

\newcommand\FullPageBG[1]{
  \AddToShipoutPictureBG*{
    \AtPageLowerLeft{
      \hspace{-2.57mm}
      \includegraphics[width=\paperwidth,height=\paperheight,keepaspectratio]{#1}
    }
  }
}

\title{Otus Supercomputer}

\author[1]{Sadaf Ehtesabi}
\author[1]{Manoar Hossain}
\author[1]{Tobias Kenter}
\author[1]{Andreas Krawinkel}
\author[1]{Holger Nitsche}
\author[1]{Lukas Ostermann}
\author[1,2]{Christian Plessl}
\author[1]{Heinrich Riebler}
\author[1]{Stefan Rohde}
\author[1]{Robert Schade}
\author[1]{Michael Schwarz}
\author[1]{Jens Simon}
\author[1]{Nils Winnwa}
\author[1]{Alex Wiens}
\author[1]{Xin Wu}

\affil[1]{Paderborn Center for Parallel Computing (PC2), Paderborn University, Germany}
\affil[2]{Department of Computer Science, Paderborn University, Germany}

\affil{\texttt{firstname.lastname@uni-paderborn.de}}

\publishers{Version 1.0}

\begin{document}

\begin{titlepage}
  \FullPageBG{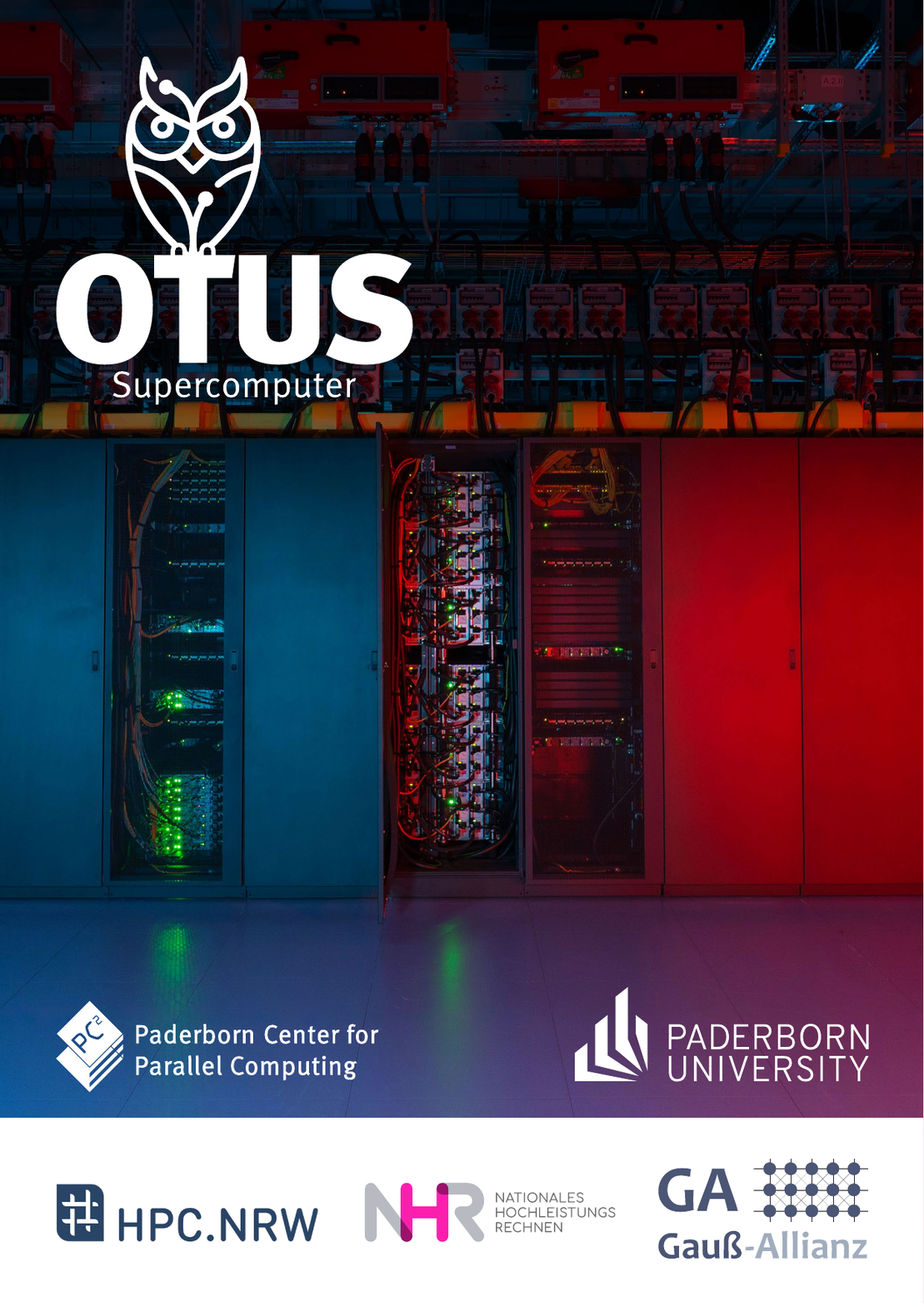}
  \null\vfill
  \vfill\null
\end{titlepage}

\ClearShipoutPictureBG

\maketitle

\begin{abstract}
\OTUS is a high-performance computing cluster that was launched in $2025$ and is operated by the Paderborn Center for Parallel Computing (\PCTWO) at Paderborn University in Germany. The system is part of the National High Performance Computing (NHR) initiative. \OTUS complements the previous supercomputer \NOCTUATWO, offering approximately twice the computing power while retaining the three node types that were characteristic of \NOCTUATWO: 1) CPU compute nodes with different memory capacities, 2) high-end GPU nodes, and 3) HPC-grade FPGA nodes. On the Top500 list, which ranks the $500$ most powerful supercomputers in the world, \OTUS is in position $164$ with the CPU partition and in position $255$ with the GPU partition (June $2025$). On the Green500 list, ranking the 500 most energy-efficient supercomputers in the world, \OTUS is in position $5$ with the GPU partition (June $2025$). 

This article provides a comprehensive overview of the system in terms of its hardware, software, system integration, and its overall integration into the data center building to ensure energy-efficient operation. The article aims to provide unique insights for scientists using the system and for other centers operating HPC clusters. The article will be continuously updated to reflect the latest system setup and measurements. 
\end{abstract}

\clearpage
\tableofcontents

\newpage
\section{Introduction}

High-performance computing (HPC) powers modern society and research. HPC clusters must provide an extensive amount of computing resources, including various accelerators such as graphics processing units (GPUs) and field-programmable gate arrays (FPGAs), in order to support a variety of workloads. User applications require a massive amount of fast, parallel storage, as well as high-throughput, low-latency networking to scale workloads over hundreds or thousands of cores. At the same time, the need for efficient, service-oriented operation is also increasing. Data center buildings need to provide reliable power, effective cooling, and an elaborate layout for maintenance and monitoring. 

\begin{figure}[h!]
\centering
\includegraphics[width=1\textwidth]{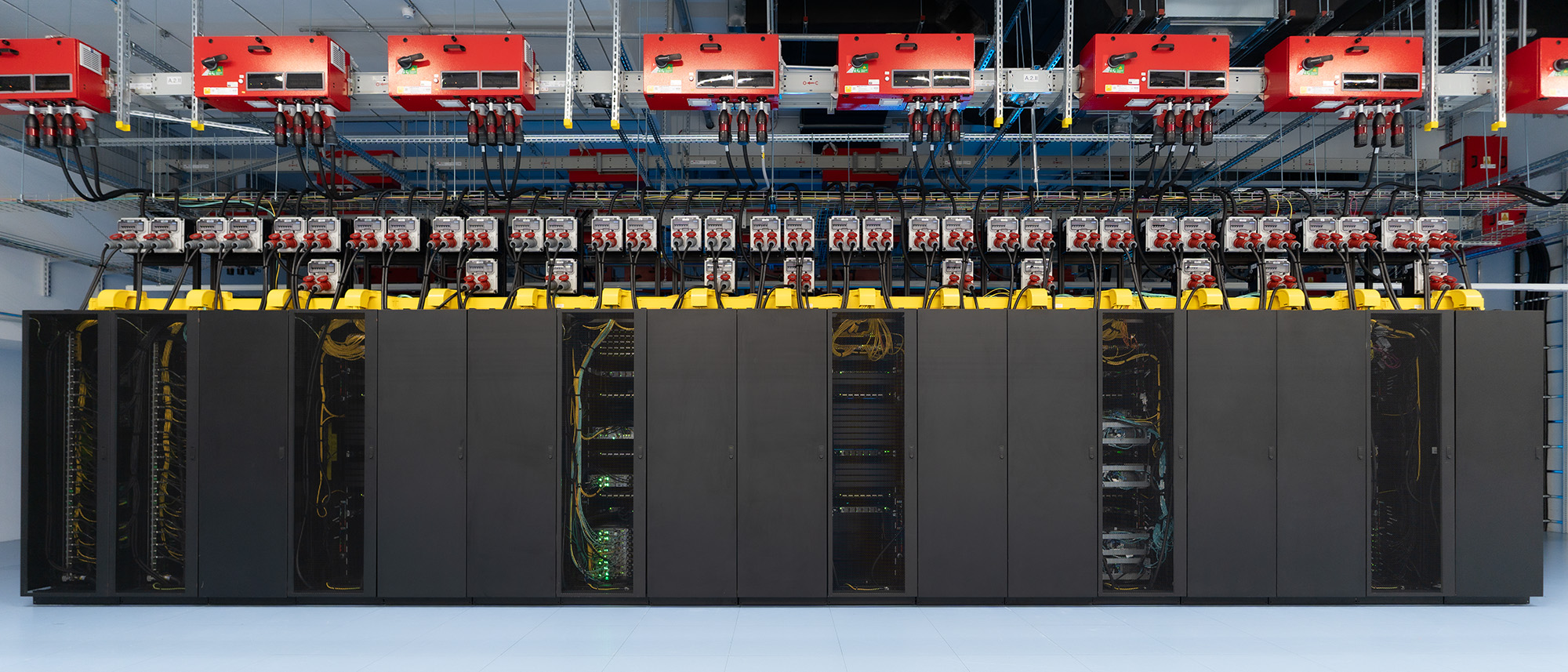}
\caption{
\OTUS supercomputer operated at the Paderborn Center for Parallel Computing.}
\label{fig:otus_real}
\end{figure}

The Paderborn Center for Parallel Computing  \citep{PC2} is a scientific institute at Paderborn University in Germany. Its mission is to advance interdisciplinary research in parallel computing and computational sciences using novel methods and innovative high-performance computer systems. As a national HPC computing center, \PCTWO collaborates in the National High Performance Computing \citep{NHR} initiative, the Gauß Alliance \citep{GA}, and the regional \citep{HPCNRW} competence network of North Rhine-Westphalia with the goal of providing HPC infrastructure and services for academic users. \PCTWO has three main focus areas: atomistic simulations, optoelectronics and quantum photonics, and machine learning for intelligent systems. Users benefit from the long-standing expertise in these areas and highly qualified personnel. 

\OTUS (see \autoref{fig:otus_real}) is the current flagship HPC system operated at \PCTWO. It was inaugurated in November 2025 and is the main focus of this article. It is currently operated in parallel to \NOCTUATWO \citep{bauer2024noctua}. Access to the systems is science-guided, peer-reviewed, and competitive. We serve academic users from public or state-recognized, institutionally accredited German universities (including FH/HAW), as well as FPGA developers worldwide. Further information is provided on the system access page \citep{PC2Access}. \\

The article is outlined as follows: \autoref{sec:overview-otus-hpc-cluster} gives an overview of the overall system components. \autoref{sec:interconnect:otus} describes the system interconnect and network topology, while \autoref{sec:overview-otus-hpc-cluster:parallel-file-system} describes the storage subsystem. Subsequently, \autoref{sec:cpu-nodes}, \autoref{sec:gpu-nodes}, and \autoref{sec:fpga-infrastructure} describe the three node types (CPUs, GPUs, and FPGAs), providing detailed measurements. In \autoref{otus-software-and-system-management} the software stacks, services, and system management are presented. \autoref{sec:noc2:power-and-cooling} shows the overall data center view, focusing on the technical infrastructure required to operate high-performance clusters reliably and efficiently. Finally, \autoref{sec:conclusion} concludes the article, and the \hyperref[sec:appendix]{Appendix} provides supplementary material for interested readers, such as visuals of the system setup.

\section{Cluster Overview}
\label{sec:overview-otus-hpc-cluster}
This section introduces the overall architecture of the \OTUS cluster. The computing and storage components were manufactured by Lenovo, while the installation and maintenance are handled by pro-com Datensysteme GmbH. 

\begin{figure}[h!]
\makebox[\textwidth][c]{\includegraphics[width=1.1\textwidth]{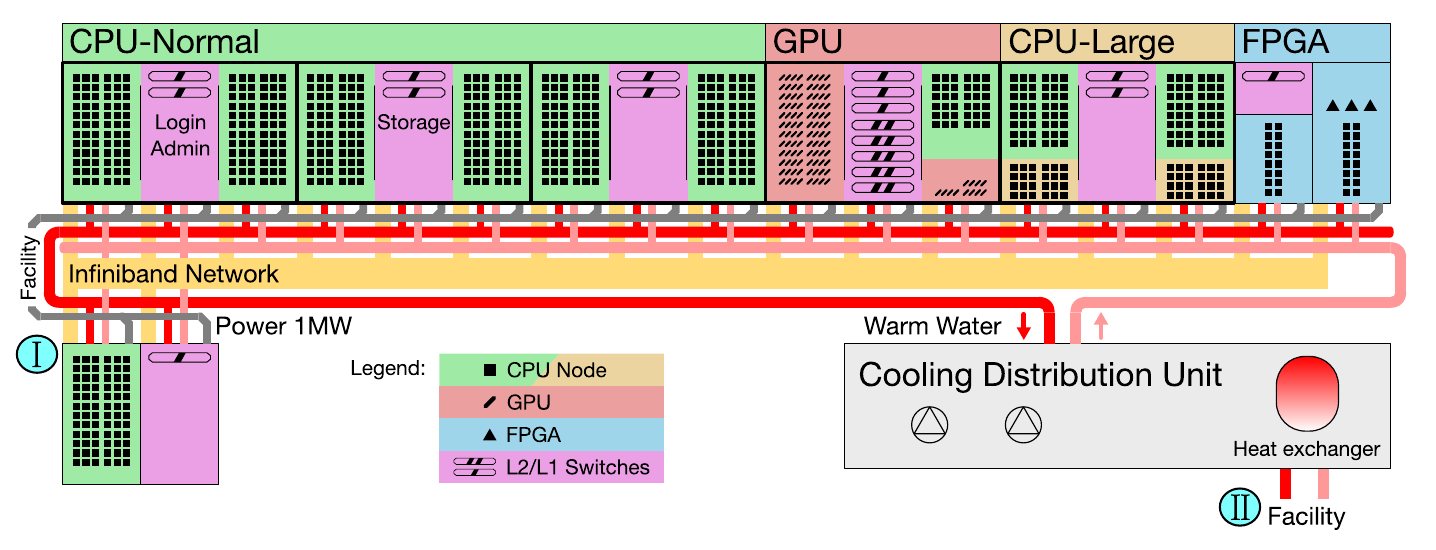}}
\caption[]{
Schematic floor plan of \OTUS, showing the racks containing nodes, network switches, and accelerators, along with the power \circledcyan{\uproman{I}} and water cooling infrastructure.
Three racks are grouped into a \emph{pod} with the middle rack hosting management and networking devices.
Single CPU nodes of \emph{normal} and \emph{largemem} variants are represented by a square.
Individual GPU and FPGA accelerators are represented, respectively.
The InfiniBand network connects all devices, including the storage.
The cooling distribution unit pumps warm water to the racks and returns heated water.
A heat exchanger exchanges the heat between the cluster water loop and the facility water loop \circledcyan{\uproman{II}}. }

\label{fig:otus_floor}
\end{figure}

\autoref{fig:otus_floor} provides a schematic floor plan showing the racks containing the nodes, together with the infrastructure required to power and to cool the system. The top row consists of five \textit{pods} (each containing three racks), as well as two additional FPGA racks on the far right. Within each pod, the middle rack houses the management nodes, storage, and networking hardware, while the two outer racks contain the corresponding compute nodes (CPU-normal, GPU, or CPU-largemem). Two further racks, which serve as expansion capacity, are located in the bottom row and integrate into the common power, network, and water-cooling pathways.
All racks are connected to the central Cooling Distribution Unit (CDU), which pumps the liquid for the rack water cooling loop. The CDU is connected via a heat exchanger to the water loop of the computing center facility. The CDU occupies the floor space of seven racks. Appendix~\ref{sec:appendix-visuals} shows images of \OTUS from different perspectives and of the CDU. It also illustrates power distribution, InfiniBand networking, and the water-cooling loop of the computing center facility. The blue circles (\circledcyan{\uproman{I}} and \circledcyan{\uproman{II}}) indicate how \OTUS is connected to the facility's power delivery and warm water loop. These connections are explained in more detail in \autoref{sec:noc2:power-and-cooling}. 

With the existing CDU capacity, the cluster can be extended by one additional compute rack of any type. To facilitate more racks with cooling, another CDU is required. 
Unlike the \NOCTUATWO cluster, where every rack's heat exchanger connects the rack cooling loop to the facility, the \OTUS cluster structure is more centralized, with the central heat exchanger operated in the CDU. Initial observations show an efficiency advantage over the \NOCTUATWO cluster's decentralized cooling solution at roughly the same thermal output. Once sufficient operational data is available, these observations will be quantified through long-term measurements in a future version of the article.

\newpage
\subsection*{Node Types and Quantities}
\label{cluster-overview:node-types}

Similar to \NOCTUATWO, the \OTUS cluster partitions are differentiated by the size of the node's main memory and the types of accelerator.
\autoref{tab:partitions} shows the different partitions and resources.

\begin{figure}
\centering
\includegraphics[width=1.0\textwidth]{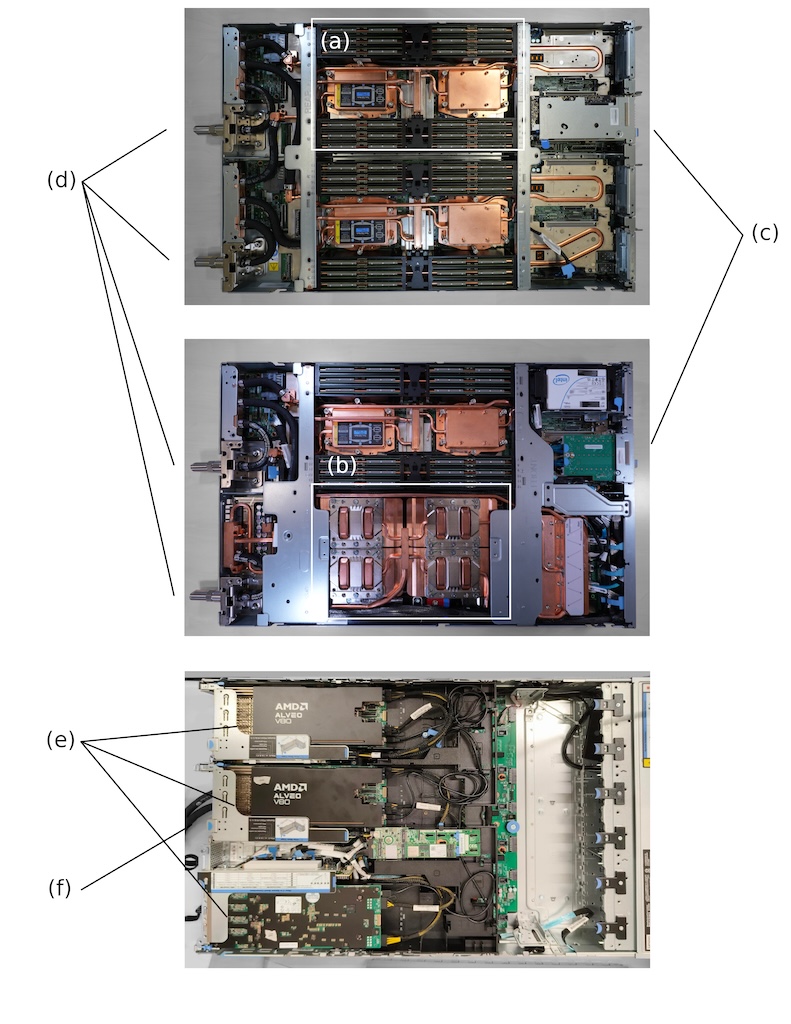}
\caption{
Blades of different partitions on \OTUS.
Top: blade from \textit{normal} partition.
Middle: blade from \textit{gpu} partition.
Bottom: 2 rack unit blade from \textit{fpga} partition.
(a) 1 node with 2 CPU sockets and main memory.
(b) 4 NVIDIA H100 GPUs in a GPU node.
(c) InfiniBand interconnect adapters.
(d) direct water cooling connectors.
(e) 3 FPGA PCIe cards.
(f) open loop water cooling inlet/outlet.
}
\label{fig:otus_blades}
\end{figure}

In total, \OTUS has 743 nodes.
All nodes have two AMD EPYC 9655 CPU sockets.
The CPU nodes in the \textit{normal} and \textit{largemem} partitions do not have accelerators and are built as two nodes per one rack unit blade.
The GPU nodes have four NVIDIA H100 GPUs and occupy one rack unit blade each.
Each FPGA node takes up two rack units and can host up to three PCIe extension cards.
\autoref{fig:otus_blades} shows all three types of blades from a top perspective.

\begin{table}[h!]
\centering
\begin{tabular}{|c|c|c|c|c|c|}
\multicolumn{6}{l}{\textbf{CPU Nodes}} \\ \hline
\myrowcolourCPU
Partition    & Nodes & CPU & Memory & InfiniBand & Specialty \\ \hline
normal &  636  & \multirow{2}{*}{EPYC 9655} & \SI{768}{\gibi\byte} & \multirow{2}{*}{NDR200} & - \\ \cline{1-2}\cline{4-4} \cline{6-6}
largemem  &   48  &                            & \SI{1.5}{\tebi\byte} &                         & local $1\times$ \SI{3.8}{\tera\byte} NVMe SSD                   \\ \hline
\multicolumn{6}{l}{} \\
\multicolumn{6}{l}{\textbf{GPU Nodes}} \\
\myrowcolourGPU
\hline
Partition  & Nodes & CPU       & Memory               & InfiniBand & Accelerator                        \\ \hline
\multirow{2}{*}{gpu} &    \multirow{2}{*}{27} & \multirow{2}{*}{EPYC 9655} & \multirow{2}{*}{\SI{768}{\gibi\byte}} & \multirow{2}{*}{$2\times$ NDR400}  & $4\times$ NVIDIA H100 \SI{94}{\giga\byte} \\
&&&&& local $1\times$ \SI{3.8}{\tera\byte} NVMe SSD  \\
\hline
\multicolumn{6}{l}{} \\
\multicolumn{6}{l}{\textbf{FPGA Nodes}} \\
\myrowcolourFPGA
\hline
Partition & Nodes & CPU       & Memory                                & InfiniBand & Accelerator                      \\ \hline
fpga    &   32* & EPYC 9655 & \multirow{1}{*}{\SI{768}{\gibi\byte}} & NDR200     & total of 3x Xilinx Alveo V80     \\ \hline
\multicolumn{6}{r}{\small{*$32$ is the overall reserved size. $3$ nodes are equipped with FPGAs in a pilot phase.}} \\
\end{tabular}
\caption{
Overview of node types and quantities available in \OTUS.  
}
\label{tab:partitions}
\end{table}

\section{System Interconnect and Network Topology}
\label{sec:interconnect:otus}

The \OTUS interconnect is designed similarly to the \NOCTUATWO interconnect as a two-level fat tree, but uses upgraded technology.
Once again, an Ethernet network is employed for node management and connection to other networks.
All management and compute nodes, as well as the cluster storage system, are also connected via InfiniBand for intra-cluster communication.
The InfiniBand network in \OTUS uses the \textit{Next Data Rate} (NDR) specification, as opposed to the \textit{High Data Rate} (HDR) used in \NOCTUATWO.
These specifications define the hardware and software implementation and determine the achievable performance.
In \OTUS, the InfiniBand network was upgraded to a newer specification to meet the growing demands of HPC applications for low-latency and high-bandwidth data transfer.

\begin{figure}[h!]
\centering
\includegraphics[width=0.7\textwidth]{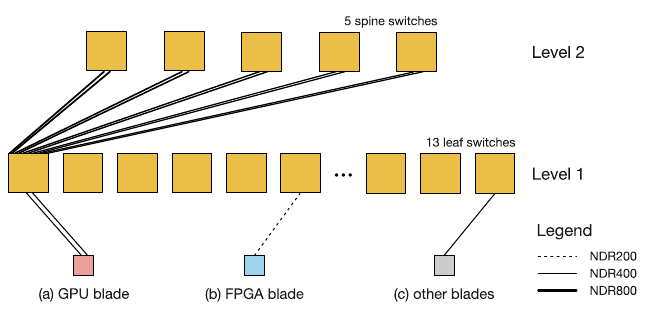}
\caption{
InfiniBand network topology used in \OTUS. Every level 1 leaf switch is connected to all five level 2 spine switches. The (a) GPU, (b) FPGA and (c) other compute blades are connected to the leaf switches.}
\label{fig:otus_network}
\end{figure}

\newpage
\autoref{fig:otus_network} depicts the InfiniBand fat-tree topology, comprising five \textit{level 2} spine switches and thirteen \textit{level 1} leaf switches. The leaf switches are connected via $2\times$ NDR800 to the spine switches, and via $2\times$ NDR400, $1\times$ NDR200, and $1\times$ NDR400 to the GPU blades (a), FPGA blades (b), and other compute blades (c), respectively.  
Since two CPU nodes are hosted on a blade, each node is connected at speeds of at least \SI{200}{\giga\bit\per\second}.
Both the leaf and spine switches use 32-port NDR800 InfiniBand: each leaf switch connects to the spine switches via 10 ports and to its compute nodes via up to 22 ports.
This topology results in a blocking factor of 1:2.
Measurements show a latency of around \SI{1.2}{\micro\second} in an MPI ping-pong test and a bidirectional bandwidth of around \SI{48}{\giga\byte\per\second}.

\begin{figure}[h!]
\centering
\includegraphics[width=0.8\textwidth]{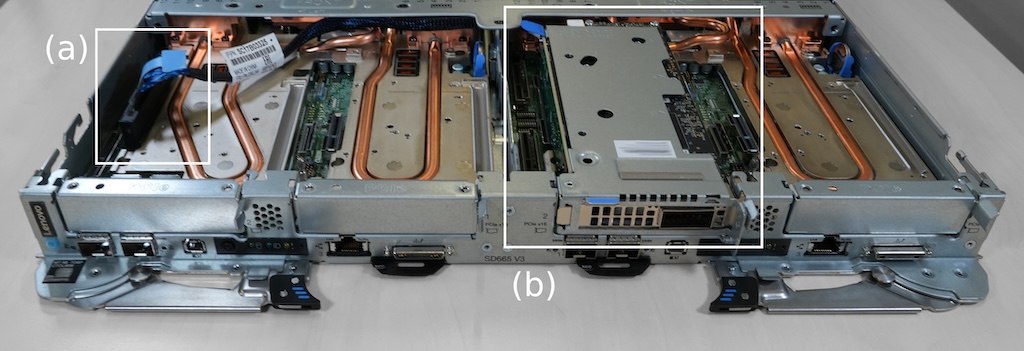}
\caption{
I/O side of a CPU compute blade. (a) SharedIO PCIe connector connects the InfiniBand adapter on the right to the PCIe slot of the left compute node.
(b) InfiniBand card in the right compute node's PCIe slot.
}
\label{fig:cpu_blade_back}
\end{figure}

The \OTUS floor plan in \autoref{fig:otus_floor} shows the distribution of InfiniBand switches across the cluster's racks.
\autoref{fig:cpu_blade_back} depicts the I/O side of a CPU node in the \textit{normal} partition with two nodes per blade.
The \textit{SharedIO} feature enables both nodes to connect to the InfiniBand network using just one adapter and one leaf switch port.

\newpage
\section{Storage Subsystem}
\label{sec:overview-otus-hpc-cluster:parallel-file-system}

\OTUS offers a variety of file systems designed for specific purposes. \autoref{tab:storage:overview} provides an overview and details the key differences. The overall structure is almost identical to that of \NOCTUATWO. All file systems have hard quotas enabled. Once these limits are reached, no more data can be written. These limits are set according to the storage capacity (used blocks) and the number of files (used inodes).

\begin{table}[h!]
\centering
\begin{tabular}{|c|p{6cm}|c|c|c|c|}
\hline
\myrowcolour
 &   & \multicolumn{2}{c|}{Permission Level} & \\
\myrowcolour
Name & Purpose  & Compute Nodes & Login Nodes & Backup \\
\hline
HOME & User home directory for permanent, small data          & \multirow{3}{*}{read-write} & \multirow{5}{*}{read-write} & \multirow{2}{*}{yes}           \\
\cline{1-2}\cline{5-5}
PC2PFS & Parallel file system of \OTUS (GPFS)                 &                             &                             & no \\ 
\cline{1-3}\cline{5-5}
PC2DATA & Permanent project data for binaries and final results  & \multirow{4}{*}{read only}  &                             & \multirow{2}{*}{yes}        \\
\cline{1-2}\cline{5-5}
PC2PFSN2 & Parallel file system of \NOCTUATWO (Lustre)        &                             &                             & \multirow{2}{*}{no}        \\
\cline{1-3}\cline{5-5}
PC2DEPOT & Long-term backup of research data                  & not available               &                             & yes        \\
\hline
\end{tabular}
\caption{
Overview of the different storage types in \OTUS.}
\label{tab:storage:overview}
\end{table}

\begin{figure}
    \centering
    \includegraphics[width=0.3\linewidth]{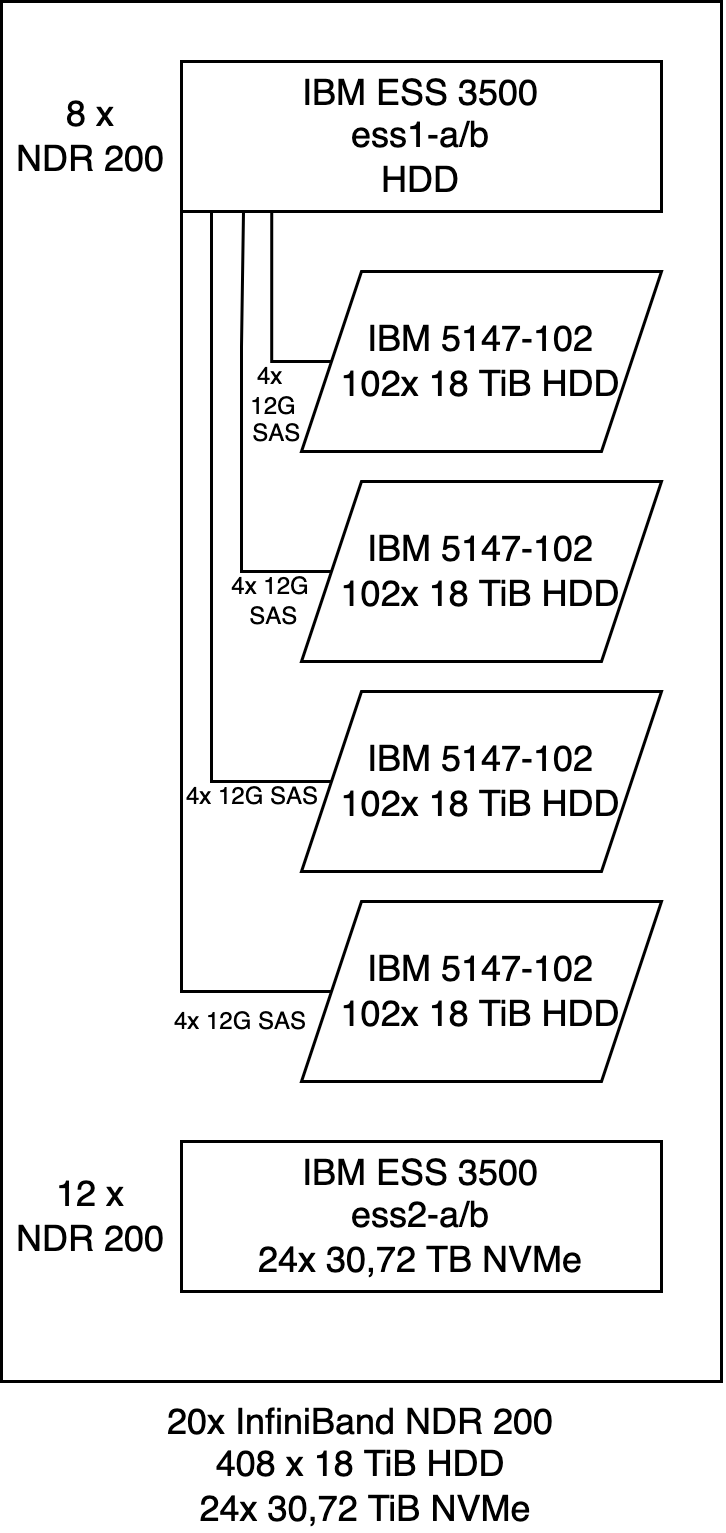}
    \caption{Illustration showing the building blocks of the GPFS.}
    \label{fig:otus-gpfs}
\end{figure}

The parallel file system is an IBM Storage Scale file system (formerly IBM Spectrum Scale and GPFS; see \citep{gpfs}). It has a capacity of \SI{5}{\peta\byte}. As depicted in \autoref{fig:otus-gpfs}, the file system is built using two IBM ESS 3500 servers, each of which contains two \textit{canisters} for high availability. One ESS server provides the Flash Pool with $24\times$ \SI{30,72}{\tera\byte} NVMe SSDs. The other ESS server handles the HDD Pool with four IBM JBODs, each containing $102\times$ drives with a capacity of \SI{18}{\tera\byte}. 

The IO500 benchmark \citep{io500} was performed according to the ISC23 specification\footnote{\url{https://github.com/IO500/io500/tree/io500-isc23}}, and the results can be found in \autoref{tab:io500}.

\begin{table}
\centering
\begin{tabular}{|c|rl|}
\hline
\myrowcolour
Metric & \multicolumn{2}{c|}{Performance} \\
\hline
ior-easy-write     & \SI{54.6}{}   & \SI{}{\giga\byte\per\second} \\
\myrowcolourB
mdtest-easy-write  & \SI{268.6}{}  & \SI{}{\kilo\iops} \\
mdtest-hard-write  & \SI{43.0}{}   & \SI{}{\kilo\iops} \\
\myrowcolourB
ior-easy-read      & \SI{107.2}{}   & \SI{}{\giga\byte\per\second} \\
mdtest-easy-stat   & \SI{356.4}{}  & \SI{}{\kilo\iops} \\
\myrowcolourB
mdtest-hard-stat   & \SI{353.6}{}  & \SI{}{\kilo\iops} \\
mdtest-easy-delete & \SI{179,2}{}  & \SI{}{\kilo\iops} \\
\myrowcolourB
mdtest-hard-read   & \SI{346,4}{}  & \SI{}{\kilo\iops} \\
IOR-Single-Node Read & \SI{28.27}{}   & \SI{}{\giga\byte\per\second} \\
\myrowcolourB
IOR-Single-Node Write & \SI{25.51}{}   & \SI{}{\giga\byte\per\second} \\
\hline
\end{tabular}
\caption{
Selected results of the IO500 benchmark performed on \OTUS according to the ISC23 specification.}
\label{tab:io500}
\end{table}

\noindent Due to the mixed performance of the parallel file system (PFS) of \NOCTUATWO, particularly with small files, a different setup was chosen for the \OTUS cluster. The IBM Storage Scale parallel file system was selected, primarily due to the file placement policies available  with ILM~\citep{ibmilm} (such as first write on NVMe). Thanks to these policies and the additional  metadata resources, the expected performance and interactivity should avoid the issues experienced with \NOCTUATWO.

\section{CPU Nodes}
\label{sec:cpu-nodes}

The CPU compute nodes in \OTUS contain only CPU resources and no accelerator cards.
The \textit{normal} partition contains 636 compute nodes with \SI{768}{\gibi\byte} of main memory, and the \textit{largemem} partition contains 48 compute nodes with \SI{1.5}{\tebi\byte} of main memory.
The \textit{largemem} and \textit{gpu} nodes also include local storage of $1\times$ \SI{3.8}{\tera\byte} NVMe SSD for temporary storage during computations, which are automatically managed by the job scheduler.
The following sections describe details of the CPU architecture in detail and present the results of selected performance measurements on the CPU nodes.

\begin{figure}[h!]
\centering
\includegraphics[width=1\textwidth]{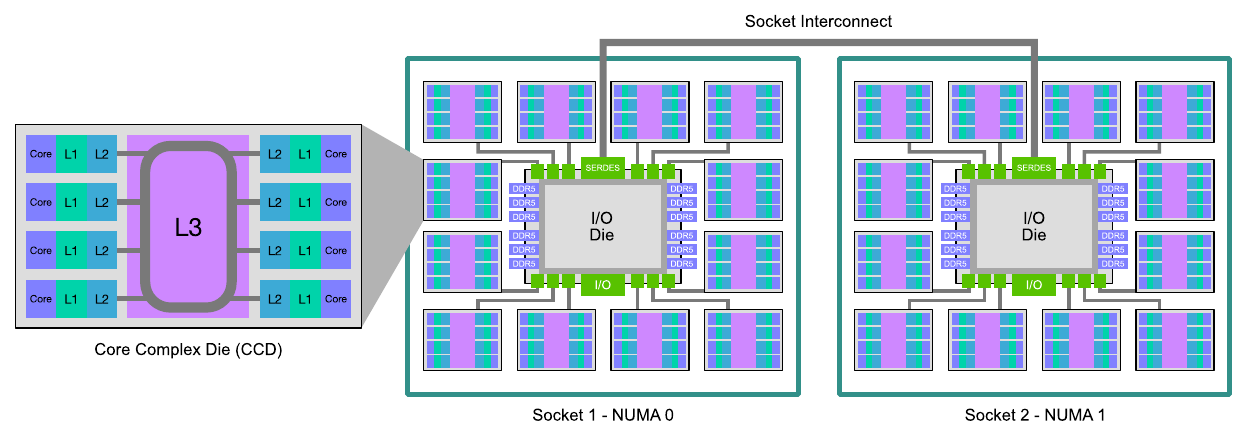}
\caption{
Topology of two AMD EPYC Turin CPU sockets.
Each CPU has 12 dies with 8 CPU cores each.
The central I/O die handles connectivity with main memory, I/O devices and the other CPU socket.
}
\label{fig:turin_cpu}
\end{figure}

\subsection{CPU and Memory Hierarchy}

\OTUS nodes feature the latest server-grade CPUs from AMD.
The AMD EPYC Turin series CPU models contain Zen 5 or Zen 5c cores.
The CPU model used in \OTUS, the EPYC 9655, has 96 Zen 5 cores spread over 12 chiplet dies, with 8 cores per CCD/CCX (Core-Cache-Die/Core-Cache-Complex).
The CPU package also contains an I/O die that connects all the core dies and handles communication between the cores, the main memory, the other socket, and PCIe devices.
Each \OTUS node hosts two sockets with EPYC 9655 CPUs, see \autoref{fig:turin_cpu}.
This results in a total of 192 cores per node in \OTUS.

The Zen 5 architecture supports the AVX-512 (advanced vector extensions) instruction set extension for x86 CPUs, extending the vector registers to 512 bits and adding mask registers, among other useful features. In contrast, the AMD EPYC Milan CPUs in the \NOCTUATWO cluster feature 8 CCDs with a total of 64 cores (128 cores per node) and only support AVX2.

Another important difference is the clock frequency and power usage.
The EPYC 9655 used in \OTUS clocks up to \SI{4.5}{\giga\hertz} (\SI{2.6}{\giga\hertz} base frequency) and has a thermal design power (TDP) of \SI{400}{\watt} per socket. In contrast, the EPYC 7763 in \NOCTUATWO clocks up to \SI{3.5}{\giga\hertz} (\SI{2.0}{\giga\hertz} base frequency) with a TDP of \SI{280}{\watt}. \\

Similar to \NOCTUATWO, simultaneous multithreading (SMT) is disabled on compute nodes of \OTUS because of the negligible gain in typical HPC workloads.

\autoref{tab:cpu_caches} lists the cache hierarchy of the EPYC 7763 and EPYC 9655 CPUs.
In comparison, the Zen 5 architecture of the 9655 provides a larger level 1 data cache and level 2 cache.
As will be seen in the following section, the level 3 cache latency has also been improved.

\begin{table}[h!]
\centering
\begin{tabular}{|c|c|c|c|c|}
    \multicolumn{4}{l}{\textbf{EPYC 7763/7713 Caches}} \\
    \hline
    \myrowcolourCPU
    Type     & Associativity   & Size               & Latency & Meas. Latency \\ \hline
    L1 Instr &  \multirow{4}{*}{8-way set associative} & \multirow{3}{*}{\SI{32}{Ki\byte} per core} & - & - \\
    \cline{1-1}\cline{4-5}
    \multirow{2}{*}{L1 Data}  &                                  &                                & 4-5 cycles (Integer) & \multirow{2}{*}{$\sim$ 4 cycles} \\

    &&& 7-8 cycles (FPU) & \\
        \cline{1-1}\cline{3-5}
    L2       &                                  & \SI{512}{Ki\byte} per core     & 12 cycles & $\sim$ 7-20 cycles \\
    \hline
    L3       & 16-way set associative & \SI{32}{Mi\byte} per 8 cores   & 46 cycles & $\sim$ 24-100 cycles \\ \hline
    \multicolumn{4}{c}{} \\
    \multicolumn{4}{l}{\textbf{EPYC 9655 Caches}} \\
    \myrowcolourCPU
    \hline
    Type     & Associativity   & Size               & Latency & Meas. Latency \\ \hline
    L1 Instr &  8-way set associative & \SI{32}{Ki\byte} per core      & - & - \\
    \hline %
    L1 Data & 12-way set associative  & \SI{48}{Ki\byte} per core       & - & $\sim$ 4 cycles \\
    \hline %
    L2   &  16-way set associative & \SI{1024}{Ki\byte} per core      & - & $\sim$ 7-18 cycles \\
    \hline
    L3   & 16-way set associative & \SI{32}{Mi\byte} per 8 cores & - & $\sim$ 23-50 cycles    \\ \hline
\end{tabular}
\caption{
Cache hierarchy of AMD EPYC processors \citep{devices2020software,amd2025hotchips} used in \NOCTUATWO (EPYC 7763/7713) and \OTUS (EPYC 9655) listing details of level 1 (L1) to level 3 (L3) CPU caches. See \autoref{fig:cache_latency_comparison_cycles} for measured latency values.
}
\label{tab:cpu_caches}
\end{table}

\begin{figure}[h!]
\centering
\includegraphics[width=0.6\textwidth]{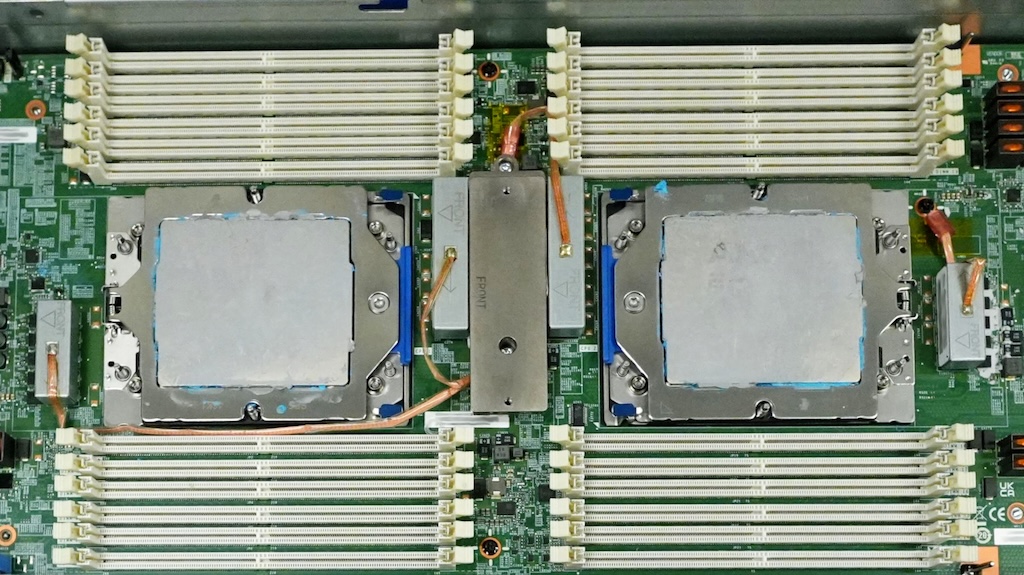}
\caption{
The two AMD EPYC CPUs are visible when removing the heat spreader. 12 (empty) DDR5 DIMM slots are located next to each CPU.}
\label{fig:cpu_sockets}
\end{figure}

Each CPU has 12 channels for DDR5 memory.
\autoref{fig:cpu_sockets} shows a node from the \textit{normal} partition during maintenance.
The heat spreader has been removed, revealing the two CPUs with thermal paste in their respective sockets.
There are 24 slots for DDR5 dual inline memory modules (DIMMs) next to them, with 12 slots for each CPU.
The \OTUS nodes use DDR5-6400 DIMMs.

The combination of all of these differences results in considerable performance improvements, as detailed in the subsequent section.

\newpage
\subsection{Theoretical and Measured CPU Performance}
The following sections examine the theoretical and measured performance of the CPUs in \OTUS.
Next to synthetic computational and data transfer attributes, the performance of selected applications is also investigated.

\paragraph{Synthetic floating-point performance}
Theoretical CPU performance can be estimated by assuming that a certain number of instructions executing a certain number of floating-point operations can be issued in each clock cycle.
Due to the pipelined architecture of the CPU and execution units, the number of instructions issued per cycle can be determined.
Depending on the availability of sufficient free and compatible execution units, the CPU can dispatch multiple instructions in one cycle.
Vector instructions perform operations on multiple elements within a single instruction.
Additionally, fused multiply-and-add (FMA) instructions perform two operations per element.
Taking all these factors into account, the performance can be estimated as follows:
\begin{flalign*}
    && \text{sockets} \times \text{cores} \times \text{frequency} \times  \text{execution units} \times \text{elements per vector} \times \text{operations per element}
\end{flalign*}
If other instructions can be issued in the same cycle as the FMA instructions, then the theoretical performance can be increased further. It is important to note that this assumed mix of instructions $-$ FMA instructions without further memory operations $-$ is mostly synthetic and serves to demonstrate the technically achievable performance limits. \\

\autoref{tab:cpu_estimated_flops} lists estimations for the relevant double-precision variants on AMD EPYC 7763 (used in the \NOCTUATWO cluster) and AMD EPYC 9655 (used in the \OTUS cluster).

\begin{table}[h!]
\centering
\begin{tabular}{|c|c|c|c|c|c|r|rl|}
\myrowcolourCPU
\hline
CPU &  Cores & Freq. & Exec. & Ele. & Op. & Performance & \multicolumn{2}{c|}{Variant}\\
\myrowcolourCPU
&& [\SI{}{\giga\hertz}] & Units &&& [\SI{}{\tera\flop/\second}] &&  \\ \hline
AMD 7763                  & \multirow{2}{*}{128} & \multirow{2}{*}{2.45} & 2 & 4 & 2 & 5.017  & AVX2    & FMA       \\
(\NOCTUATWO)              &                      &                       & 2 & 4 & 3 & 7.526  & AVX2    & FMA+ADD   \\  \cline{1-9}
                          & \multirow{4}{*}{192} & \multirow{4}{*}{2.60} & 2 & 4 & 2 & 7.987  & AVX2    & FMA       \\
AMD 9655                  &                      &                       & 2 & 4 & 3 & 11.980 & AVX2    & FMA+ADD   \\
\multirow{1}{*}{(\OTUS)}  &                      &                       & 2 & 8 & 2 & 15.974 & AVX-512 & FMA       \\
                          &                      &                       & 2 & 8 & 3 & 23.961 & AVX-512 & FMA+ADD   \\ \hline
\end{tabular}
\caption{Theoretical CPU performance estimation on \NOCTUATWO and \OTUS for double-precision floating-point operations.}
\label{tab:cpu_estimated_flops}
\end{table}

The estimated variants were measured using \textit{likwid-bench} \citep{treibig2010likwid}, which executes a microbenchmark in the form of a user-defined assembly instruction sequence in a loop.
\autoref{tab:cpu_synthetic_measured_flops} lists the floating-point performance results and approximate CPU clock readings.
Clearly, the increased number of cores, AVX-512 support and higher CPU clock result in significantly higher performance on \OTUS.
Note that, on both systems, the observed CPU frequency under full load is higher than the documented base frequency used in the estimation, and varies depending on the workload.

\begin{table}[h!]
\centering
\begin{tabular}{|c|c|c|r|rl|}
\myrowcolourCPU
\hline
System  & Cores & Frequency [\SI{}{\giga\hertz}] & Performance [\SI{}{\tera\flop/\second}] & \multicolumn{2}{c|}{Variant} \\
\hline
\multirow{2}{*}{\NOCTUATWO} & \multirow{2}{*}{128} & 2.692 & 5.449  & AVX2    & FMA       \\
                            &                      & 2.498 & 6.793  & AVX2    & FMA+ADD   \\  \cline{1-6}
\multirow{4}{*}{\OTUS}      & \multirow{4}{*}{192} & 3.960 & 12.073 & AVX2    & FMA       \\
                            &                      & 3.620 & 15.300 & AVX2    & FMA+ADD   \\
                            &                      & 3.510 & 21.275 & AVX-512 & FMA       \\
                            &                      & 3.040 & 25.760 & AVX-512 & FMA+ADD   \\ \hline
\end{tabular}
\caption{Measured floating point performance results of synthetic double-precision microbenchmark variants on the CPUs in \NOCTUATWO and \OTUS.}
\label{tab:cpu_synthetic_measured_flops}
\end{table}

\paragraph{Cache}

The cache hierarchy is an important part of the CPU architecture since the performance of memory operations is a limiting factor for many applications.
Two interesting cache properties are cache load latency and cache bandwidth, which were measured on the \NOCTUATWO and \OTUS clusters.

\begin{figure}[h!]
\centering
\includegraphics[width=1\textwidth]{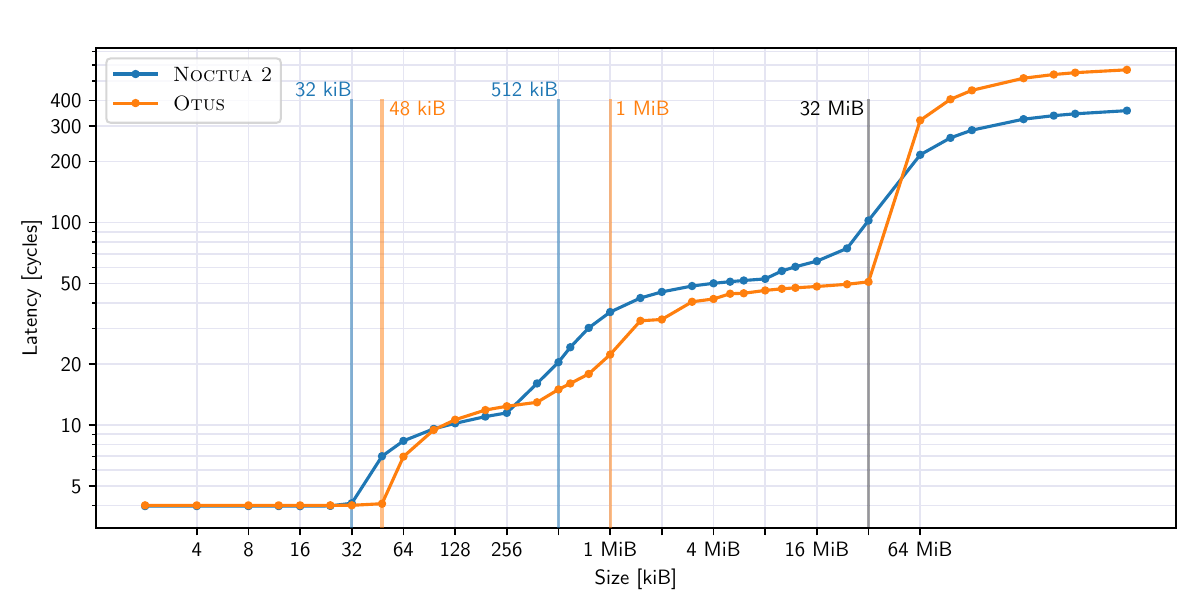}
\caption{The cache and memory latency measurement, normalized to clock cycles. Differences in level 1 and level 2 cache sizes are visible, next to a slight improvement in level 3 cache latency.}
\label{fig:cache_latency_comparison_cycles}
\end{figure}

The Chips and Cheese Memory Latency Test~\citep{cnclatency} was used to measure cache latency.
Since the cache clock is typically synchronized with the core frequency, the \OTUS cluster has a lower cache latency in nanoseconds due to its higher core frequency.
Therefore, to compare the architectural differences, the measured results can be normalized to the clock cycle length, as depicted in \autoref{fig:cache_latency_comparison_cycles}.
The different cache sizes of the level 1 and level 2 caches can be seen in the array sizes where the latency increases.
The latency of the level 1 and level 2 caches is roughly similar, while there is an improvement in the level 3 cache latency.
The difference in main memory latency is tied to the different types and speeds of DDR main memory, as well as the non-uniform memory access (NUMA) configuration used in the systems under comparison.

\begin{figure}[h!]
\centering
\includegraphics[width=1\textwidth]{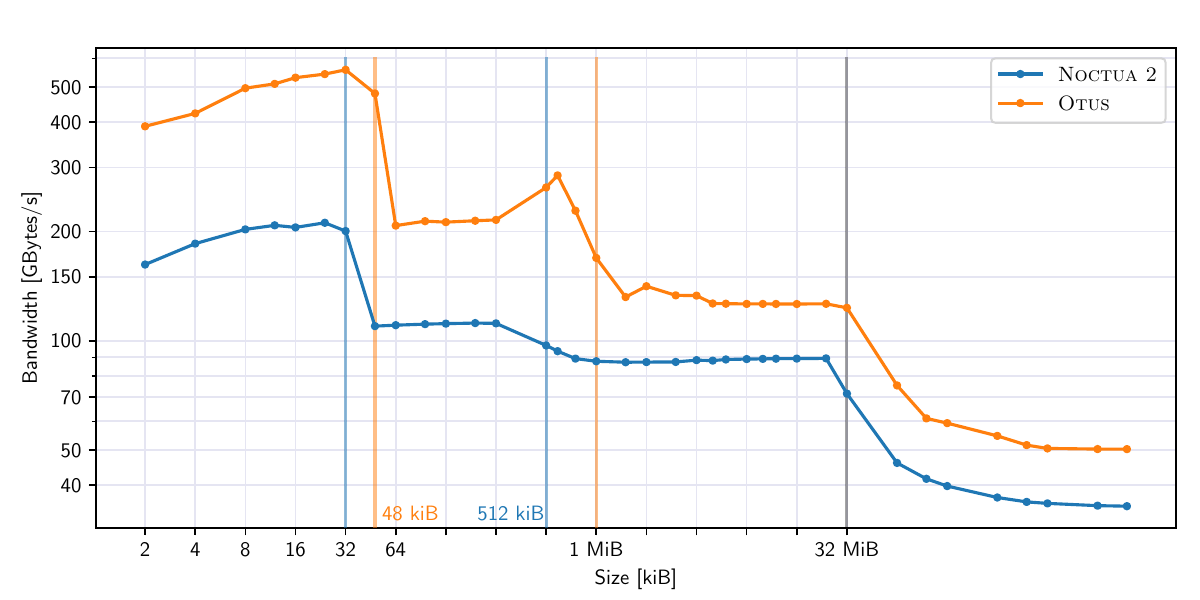}
\caption{Cache and memory bandwidth measurements on \NOCTUATWO and \OTUS using LIKWID \citep{treibig2010likwid} and one thread.
The cache bandwidth depends on the CPU core clock.}
\label{fig:cache_bandwidth_comparison}
\end{figure}

The cache bandwidth was measured using LIKWID \citep{treibig2010likwid} and is shown in \autoref{fig:cache_bandwidth_comparison}.
A clear difference in achieved bandwidth is visible.
Once again, the cache performance is tied to the core clock.

\clearpage
\paragraph{Memory and NUMA}
The theoretical memory bandwidth can be estimated using the DDR5 technical parameters:
\begin{flalign*}
   && \text{sockets} \times \text{transfers} \times \text{bytes per channel} \times \text{channels per socket}
\end{flalign*}

\OTUS has 12 memory channels per socket, in comparison to the 8 channels per socket of \NOCTUATWO.
Additionally, the \OTUS nodes support DDR5 memory, which is operated at \SI{6400}{\mega\transfer\per\second}.

\autoref{tab:cpu_gbs} lists the theoretical and measured values for main memory bandwidth on \NOCTUATWO and \OTUS.
For \OTUS, two NUMA nodes-per-socket (NPS) configurations were tested.

The NPS setting controls how data is interleaved across physical memory banks and how the operating system places memory allocations.
Higher NPS values (for example, NPS4) enable more fine-grained control over the distribution of data across memory modules.
Lower values (for example, NPS1) provide a more uniform view of the system, which simplifies programming but can slightly reduce performance.
The reported bandwidths represent the maximum achievable values.
Measurements with real applications show only small differences between the two NPS configurations, as shown in \autoref{tab:cpu_app}.
Because the performance penalty is negligible (about 1 \%), the NPS1 configuration is used on \OTUS.

\begin{table}[h!]
\centering
\begin{tabular}{|c|c|c|c|c|c|c|c|}
\myrowcolourCPU
\hline
System   & Type & NUMA & Channels & Transfer & Bytes per & Performance & Meas. Perf. \\ 
\myrowcolourCPU
&&&& rate [\SI{}{\giga\transfer\per\second}] & transfer & [\SI{}{\giga\byte\per\second}] & [\SI{}{\giga\byte\per\second}] \\ \hline
\NOCTUATWO & DDR4 & NPS4 & 16 & 3.2 & 8 &  \phantom{1}409.6 & \SI{370.3}{} \\ \hline
\OTUS     & DDR5 & NPS4 & 24 & 6.4 & 8 & 1228.8 & \SI{984.0}{} \\ \hline
\OTUS     & DDR5 & NPS1 & 24 & 6.4 & 8 & 1228.8 & \SI{970.6}{} \\ \hline
\end{tabular}
\caption{Theoretical and measured main memory bandwidth. Measurement was done using STREAM~\citep{mccalpin1995memory} on a full node. The NUMA column refers to the NUMA nodes-per-socket configuration.}
\label{tab:cpu_gbs}
\end{table}

\paragraph{Application tests}

In addition to synthetic benchmarks, it is important to conduct tests with real applications in order to review the performance of a system for certain use cases.
Due to the focus areas of the \PCTWO in the NHR alliance, especially molecular and quantum mechanical simulation applications are being executed on the computing systems.
\autoref{tab:cpu_app} shows the results of HPC-typical test benchmarks and selected applications important to the users, which are run on \NOCTUATWO and \OTUS.
A $2\text{-}3\times$ factor of performance improvement is expected due to a threefold increase in memory bandwidth and floating-point throughput. The results confirm this expectation.
High Performance Linpack (HPL)~\citep{petitet2004hpl} is a well-known linear algebra performance benchmark for HPC systems.
Another common benchmark to rank HPC systems is High Performance Conjugate Gradients (HPCG)~\citep{dongarra2013toward}.
For CP2K~\citep{kuhne2020cp2k}, a quantum chemistry and atomistic simulation software package, the \emph{H2O-512}\footnote{\url{https://github.com/cp2k/cp2k/tree/master/benchmarks/QS}} test case was used, which simulates ab initio molecular dynamics for a system of 512 water molecules over 10 time steps.
For the electronic-structure simulation suite QuantumESPRESSO~\citep{giannozzi2009quantum}, the medium test case\footnote{\url{https://repository.prace-ri.eu/git/UEABS/ueabs/-/tree/master/quantum_espresso/test_cases/medium}} from the United European Applications Benchmark Suite (UEABS)~\citep{ueabs} was used. This computes the DFT ground state of a crystal containing 443 atoms (GRIR443: 200 carbon and 243 iridium) and four k-points.

\begin{table}
\centering

\begin{tabular}{|r|r|r|r|}
\myrowcolourCPU
\hline
 &  \multicolumn{2}{c|}{Performance} &   \\
\myrowcolourCPU
Test & \multicolumn{1}{c|}{\NOCTUATWO} &\multicolumn{1}{c|}{\OTUS} & Improvement \\
 \hline
HPL (NPS4) [\SI{}{\giga\flop\per\second}]  &    \SI{4143}{} &  \SI{13270}{} & \SI{3.2}{\times} \\
HPCG  (NPS4) [\SI{}{\giga\flop\per\second}] &      \SI{63}{} &    \SI{182}{} & \SI{2.9}{\times} \\
\hline
CP2K (NPS4) [\SI{}{\second}] & \SI{674}{}  & \SI{222}{} & \SI{3,0}{\times} \\
CP2K (NPS1) [\SI{}{\second}] & $-$    & \SI{228}{} & $-$ \\
\hline
QuantumESPRESSO (NPS4) [\SI{}{\second}] & \SI{1139}{}  & \SI{398}{} & \SI{2.9}{\times} \\
\hline
\end{tabular}
\caption{Single CPU node tests on \NOCTUATWO and \OTUS.}
\label{tab:cpu_app}
\end{table}

\newpage
\section{GPU Nodes}
\label{sec:gpu-nodes}

The \OTUS \textit{gpu} partition consists of 27 nodes, each equipped with four NVIDIA H100 GPUs.
Each node contains four NVIDIA H100 SXM accelerators with 94GB of memory.
These models are based on the Hopper architecture, the successor to the Ampere architecture used in the A100 GPUs hosted in \NOCTUATWO.
There are multiple variants of the H100 GPU.
The H100 SXM 94GB in \OTUS connects to the mainboard as an SXM5 module and contains six stacks of HBM2E memory with a \SI{6144}{\bit} interface and a data rate of \SI{2.4}{\tera\byte\per\second}. 
The total memory per GPU is \SI{94}{\gibi\byte}.
One GPU blade combines two AMD Turin CPUs with four H100 GPUs, which are interconnected via NVLink.

\autoref{tab:gpu_specs} shows a comparison of the theoretical performance of the A100 and H100 GPUs. The H100 GPUs in \OTUS deliver multiple times the performance of A100 GPUs in \NOCTUATWO while using less than twice the power.

\begin{table}[h!]
\centering
\begin{tabular}{|l|r|r|l|}
\myrowcolourGPU
\hline
 &  A100 in \NOCTUATWO & H100 in \OTUS & \\ \hline
Streaming Multiprocessors               &   108 &   132 & \\
\myrowcolourB
Tensor cores     &   432 &   528 & \\
CUDA core FP64   &   9.7 &  33.4 & TFLOP/s \\
\myrowcolourB
CUDA core FP32   &  19.5 &  66.9 & TFLOP/s \\
Tensor core FP64 &  19.5 &  33.4 & TFLOP/s \\
\myrowcolourB
Tensor core TF32 & 155.9 & 494.7 & TFLOP/s \\
Tensor core FP16 & 311.9 & 989.4 & TFLOP/s \\
\myrowcolourB
Memory           &    40 &    94 & GiB \\
Memory type      &  HBM2 & HBM2E & \\
\myrowcolourB
Memory data rate &  1555 &  2446 & GB/s \\
NVLink bandwidth &   600 &   900 & GB/s \\
\myrowcolourB
TDP              &   400 &   700 & Watt \\
\hline
\end{tabular}
\caption{Comparison of NVIDIA A100 used in \NOCTUATWO and NVIDIA H100 used in \OTUS.}
\label{tab:gpu_specs}
\end{table}

\section{FPGA Nodes}
\label{sec:fpga-infrastructure}

High-end field-programmable gate arrays (FPGAs) are PCIe-based accelerator cards that can improve the performance of HPC applications and reduce energy consumption by customizing the reconfigurable hardware to match the application requirements (a hardware/software co-design approach).  \\

\PCTWO has many years of experience operating FPGAs in high-performance computing systems. While CPUs and GPUs are well-established computing devices in HPC, deploying FPGAs introduces distinct challenges and characteristics. Building on experience with the previous multi-FPGA system \NOCTUATWO \citep{bauer2024noctua}, \OTUS is in the process of being equipped with tens of high-end FPGAs, potentially from both main vendors (AMD and Altera). The FPGA partition is currently in a pilot phase with a limited number of FPGAs, providing an opportunity to gain hands-on experience with next-generation FPGA cards and their associated tools.

\subsection{Hardware Details}
\label{sec:fpga-infrastructure:hw-details}

An FPGA partition of the same size as in \NOCTUATWO is planned, consisting of $32$ nodes. During the current pilot phase, three AMD Alveo V80 FPGA cards have been installed. \autoref{tab:fpga-cards-overview} summarizes the main hardware features of the host servers and the V80 cards and compares them with the cards installed in \NOCTUATWO. 

Unlike the host system of \NOCTUATWO, the \OTUS CPUs are water-cooled. This makes it possible to use exactly the same CPU models in the FPGA partition as in the rest of the system, which simplifies cross-partition runs and comparisons and increases flexibility for reallocating nodes between partitions. The \OTUS FPGA host nodes also provide \SI{50}{\percent} more main memory and larger, faster local storage than the FPGA host nodes in \NOCTUATWO. For the Alveo V80 accelerator cards, the following improvements have been made as a result of technological advancements:
\begin{itemize}
    \item The Alveo V80 PCIe Gen5 host interface offers twice the bandwidth of the Alveo U280 cards and four times that of the BittWare 520N cards.
    \item The Alveo V80's HBM capacity is four times higher, and its bandwidth is almost double that of the Alveo U280. Both cards complement the HBM with \SI{32}{\giga\byte} of DDR4 memory; however, the Alveo V80 provides this via a single channel, offering only two thirds of the bandwidth. The BittWare 520N card provides three times the bandwidth of the Alveo V80 via its DDR4 interface, but lacks the HBM complement.
    \item The Alveo V80 network interface offers four ports, the same number as the BittWare 520N cards, but at higher data rates. Compared to the Alveo U280 cards, there are twice as many ports, each with twice the bandwidth.
\end{itemize}
The thermal design power (TDP) specified in the Alveo V80 data sheet is lower than that of its predecessors, but this value should be interpreted cautiously. Synthetic power or stress tests on the Alveo V80 can readily exceed the TDP. In contrast, the power consumption of \textit{typical} workloads remains well below the specified TDP across all card types. \cite{hpcc_fpga_in_depth} measured an average and peak power consumption of around \SI{41}{\watt} and \SI{60}{\watt} on the Alveo U280 cards and around \SI{69}{\watt} and \SI{77}{\watt} on the BittWare 520N cards. For the V80 cards, currently only the example designs that are shipped with the VRT/AVED runtime\footnote{\url{https://github.com/Xilinx/SLASH/tree/dev/examples}} are available. Executing these examples results in a power consumption of around \SI{33}{\watt}. These examples do not reflect typical workloads but illustrate that actual power draw stays far below the TDP. Measurements for typical workloads on the V80 will be included in future versions of this article.

Looking more closely at the FPGA resources of the Alveo V80 cards compared to the devices installed in \NOCTUATWO:
\begin{itemize}
    \item The FPGAs of the Alveo V80 cards provide almost twice as many logic resources as those of the Alveo U280. When targeting full 6-input LUTs implementations, the advantage over the Stratix 10 FPGAs on the BittWare 520N cards would be about $2.8\times$. However, the Stratix 10 Adaptive Logic Modules (ALMs) provide 8 inputs and two internal full adders, enabling many mappings of two LUTs per ALM and reducing the effective logic advantage of the Alveo V80 to approximately $1.4\times$ for other workloads.
    \item The Alveo V80 cards provide only about $20\%$ more DSP blocks than the Alveo U280, but the internal DSP architecture has improved substantially. From an HPC perspective, the key enhancement is native support for single-precision floating-point multiply-accumulate in a single DSP block, avoiding additional logic, while the updated fixed-point capabilities also help implement double-precision floating point with fewer extra logic resources. The Stratix 10 architecture with its Variable Precision DSP Blocks already offers complete support for single-precision floating-point multiply-accumulate, but compared to the Bittware 520N, the Alveo V80 now contains almost twice as many DSP blocks.
    \item In terms of local memory resources, the Alveo V80 provides roughly twice the improvement of the Alveo U280, in line with the logic scaling. Compared to the Stratix 10 GX 2800, the Alveo V80 has a much larger overall on-chip memory capacity, but the Stratix 10 architecture still provides a larger number of independently addressable blocks. Note that each RAMB36 block in the Versal and Ultrascale+ architectures can operate as two independent RAMB18 parts; however, in this configuration, each part is either limited to a maximum of 18-bit read and write ports or a single 36-bit port.
\end{itemize}

In terms of device integration, both the Alveo U280 and the BittWare 520N use a traditional FPGA fabric in which at least part of the PCIe and memory controller logic resides in the programmable fabric, and all interconnects are realized using configurable routing resources. This design requires a substantial portion of the available resources for the board support package, shell, or user infrastructure logic. In contrast, the V80 is built around a Versal HBM Adaptive SoC (XCV80) with hardened network-on-chip (NoC) connections between the HBM, PCIe, and networking blocks, leaving a larger share of the programmable resources available for user kernels. 

\begin{table}
\begin{adjustwidth}{-0.5cm}{}
\centering
\begin{tabular}{|c|c|cc|}
\hline
\myrowcolourFPGA            &  \OTUS Pilot Phase & \multicolumn{2}{c|}{\NOCTUATWO} \\
\hline
Card Installation Year      & 2025 & 2022 & 2018 \\
\myrowcolour
Number of Nodes             & 3 out of 32\textsuperscript{1} & 16 & 16 \\
\hline
\multicolumn{4}{r}{\small{\textsuperscript{1} 32 nodes are available for the FPGA partition of \OTUS. Currently $3$ are equipped with FPGA cards.}} \\

\multicolumn{4}{l}{} \\

\multicolumn{4}{l}{\textbf{Host Configuration}} \\
\hline
CPUs                        & $2\times$   AMD Turin EPYC 9655 & \multicolumn{2}{c|}{$2\times$   AMD Milan 7713}  \\
\myrowcolour
Main Memory                 & \SI{768}{\gibi\byte}   & \multicolumn{2}{c|}{\SI{512}{\gibi\byte}  } \\
Local Storage               & \SI{960}{\giga\byte} NVMe & \multicolumn{2}{c|}{\SI{480}{\giga\byte} SSD} \\
\myrowcolour
        & 1x AMD/Xilinx & 3x AMD/Xilinx  & $2\times$ BittWare  \\
\myrowcolour
\multirow{-2}{*}{FPGA Cards per Node}  &  Alveo V80  & Alveo U280                 & 520N  \\
\hline

\multicolumn{4}{l}{} \\

\multicolumn{4}{l}{\textbf{Card Specifications}} \\
\hline
Host Interface\textsuperscript{1} & PCIe Gen5 x8 & PCIe Gen3 x16 & PCIe Gen3 x8  \\
\myrowcolour
Host Interface Bandwidth    & \SI{31.5}{\giga\byte\per\second}  & \SI{15.8}{\giga\byte\per\second}     & \SI{7.9}{\giga\byte\per\second} \\
\hline
       & 32 GB HBM2E & 8 GB HBM2  & -  \\
\multirow{-2}{*}{High-Bandwidth Memory}       & ($2\times$ 16 GB stacks) & ($2\times$ 4 GB stacks) &   \\
\myrowcolour
HBM Bandwidth & \SI{820}{\giga\byte\per\second} & \SI{460}{\giga\byte\per\second} & - \\
\hline
DDR Memory                  & 1x 32 GB DDR4 & $2\times$ 16 GB DDR4 & $4\times$ 8 GB DDR4 \\
\myrowcolour
DDR Bandwidth               & \SI{25.6}{\giga\byte\per\second}  & \SI{38.4}{\giga\byte\per\second} & \SI{76.8}{\giga\byte\per\second} \\
\hline
  & $4\times$ QSFP56 & $2\times$ QSFP28 & $4\times$ QSFP+  \\
\multirow{-2}{*}{Network Interfaces\textsuperscript{1}} & (\SI{200}{\giga\bit\per\second})\textsuperscript{2} &  (\SI{100}{\giga\bit\per\second}) & (\SI{40}{\giga\bit\per\second}) \\
\myrowcolour
\hline
Thermal Design Power\textsuperscript{3}    & \SI{190}{\watt} & \SI{225}{\watt} & \SI{225}{\watt}  \\
\hline
\multicolumn{4}{r}{\small{\textsuperscript{1} Effectively usable values, implemented by the FPGA shell and used in our system.}} \\
\multicolumn{4}{r}{\small{\textsuperscript{2} Each port supports $2\times$ 100 Gbit/s or $4\times$ 10/25/40/50 Gbit/s.}} \\
\multicolumn{4}{r}{\small{\textsuperscript{3} The TDP does not reflect the actual power consumption for typical workloads (see Section~\ref{sec:fpga-infrastructure:hw-details} for details).}} \\

\multicolumn{4}{l}{}                                                                                                         \\

\multicolumn{4}{l}{\textbf{FPGA Architectures and Resources}} \\
\hline
       & AMD/Xilinx  & AMD/Xilinx  & Altera/Intel  \\
\multirow{-2}{*}{FPGA Architecture}       & Versal & UltraScale+ & Stratix 10 \\
\myrowcolour
FPGA Model              & XCV80 & XCU280 & Stratix 10 GX 2800 \\
Lithography             & TSMC 7 nm    & TSMC 16 nm  & Intel 14 nm  \\
\hline
\myrowcolour
LUTs & \SI{2574208}{}\textsuperscript{1}   & \SI{1303680}{}\textsuperscript{1}     & \SI{933120}{} -- \SI{1866240}{}\textsuperscript{2}  \\
DSP blocks              & \SI{10848}{} DSP58\textsuperscript{3} & \SI{9024}{} DSP48E2\textsuperscript{4} & 5,760\textsuperscript{5} \\
\myrowcolour
RAM blocks              & \SI{3741}{} RAMB36E5  & \SI{2016}{} RAMB36E2 & \SI{11721}{} M20K \\
UltraRAM blocks         & \SI{1925}{} URAM      & \SI{960}{} URAM   & -  \\
\myrowcolour
RAM capacity [\SI{}{\kibi\bit}] & \SI{134676}{} + \SI{554400}{}  & \SI{72576}{} + \SI{276480}{} & \SI{234420}{} \\
\hline
\multicolumn{4}{r}{\small{\textsuperscript{1} Complete 6-input LUTs with 64 bit configuration memory. Can be split e.g. into $2\times$ 3-input LUTs.}} \\
\multicolumn{4}{r}{\small{\textsuperscript{2} 933120 ALMs can serve as full 6-input LUTs with 64 bit configuration memory. With overall 8 inputs}} \\
\multicolumn{4}{r}{\small{and additionally 2 integrated full adders, each ALM has more options to implement two LUTs.}} \\
\multicolumn{4}{r}{\small{\textsuperscript{3} Versal DSP58s support, among other configurations, single precision \textbf{floating point} }} \\
\multicolumn{4}{r}{\small{multiply-accumulate, or $27 \times \textbf{24} + \textbf{58}$ fixed-point multiply-accumulate.}}\\
\multicolumn{4}{r}{\small{\textsuperscript{4} Ultrascale+ DSP48E2 support, among other configurations, $27 \times \textbf{18} + \textbf{48}$ fixed-point}}\\
\multicolumn{4}{r}{\small{multiply-accumulate, but need additional logic resources for floating point support.}}\\
\multicolumn{4}{r}{\small{\textsuperscript{5} Stratix 10 Variable Precision DSP Blocks support, among other configurations, single precision }}\\
\multicolumn{4}{r}{\small{\textbf{floating point} multiply-accumulate, or two independent $18 \times 19$ fixed-point multiplications.}}

\end{tabular}
\caption{Comparison of the FPGA accelerator cards between the systems.}
\label{tab:fpga-cards-overview}
\end{adjustwidth}
\end{table}

\subsection{Theoretical and Measured FPGA Performance}

This section outlines the theoretical FPGA performance and compares it with the measured performance using xbtest \citep{xbtest} for the V80 AVED base design. xbtest provides a set of built-in (pre-canned) test cases that can be adjusted with parameters. We used the default settings unless otherwise noted. During execution, the application was pinned to the same CPU socket to which the FPGA card was attached. 

\paragraph{Memory Test} 
The theoretical performance of the in-FPGA HBM memory and the on-board DDR4 memory is evaluated first. Actual performance is then measured using the xbtest memory test. \\

\textbf{High-Bandwidth Memory (HBM2E) Bandwidth}

The Versal Adaptive SoC of the V80 card has the following architecture~\citep{versal-pg313}:
\begin{itemize}
    \item It consists of two HBM2E memory stacks.
    \item Each memory stack has 8 memory channels with a dedicated memory controller. 
    \item Each memory controller operates in pseudo-channel mode, having 2 \SI{64}{\bit} pseudo-channels, which address a dedicated region in the HBM memory space. 
\end{itemize}
This gives the intermediate result for the data transfer per transaction as:
\begin{flalign*}
    && = \text{2 memory stacks} \times \text{8 memory channels per memory stack} \\ 
    && \times \text{2 pseudo channels per memory channels} \times \text{\SI{64}{\bit} per pseudo channel}   \\ 
    && = \SI{2048}{\bit} = \SI{256}{\byte}
\end{flalign*}
The clock rate of the memory controller (HBM IP) is configured to \SI{1600}{\mega\hertz}. HBM is a double data rate (DDR) memory. Hence, the data bus toggles at twice the configured clock rate (on both the rising and falling edges of the clock). This gives the overall theoretical bandwidth as:
\begin{flalign*}
    && = \text{operating frequency} \times 2 \times \text{data transfer per transaction} \\
    && = \SI{1600}{\mega\hertz} \times 2 \times \SI{256}{\byte} = \SI{819200}{\mega\byte\per\second} \approx \SI{820}{\giga\byte\per\second}
\end{flalign*}

\textbf{DDR Bandwidth}

The V80 card has one \SI{32}{\giga\byte} DDR4 DIMM module by Micron (part MTA18ASF4G72PZ-3G2F1) \citep{Alveo-DS1013}. The theoretical bandwidth can be evaluated as:

\begin{flalign*}
    && = \text{transfer rate} \times \text{effective transfers} = \SI{3200}{\mega\transfer\per\second} \times \SI{8}{\byte} = \SI{25.6}{\giga\byte\per\second} 
\end{flalign*}

\textbf{Measurements and Setup} 

The xbtest memory test case is used to evaluate communication of the in-FPGA HBM memory and the on-board DDR4 memory in four modes:
\begin{itemize}
    \item Mode 1 (only write): the entire memory range is fully written repeatedly.
    \item Mode 2 (only read): the entire memory range is fully read repeatedly.
    \item Mode 3 (alternate write and read): the entire memory range is written fully, then read fully repeated over the test duration.
    \item Mode 4 (simultaneous write and read): the entire memory range is split into two halves. At the same time and repeatedly, the first half is fully written, while the second half is fully read from.
\end{itemize}

\begin{figure}
\centering
\includegraphics[width=1\textwidth]{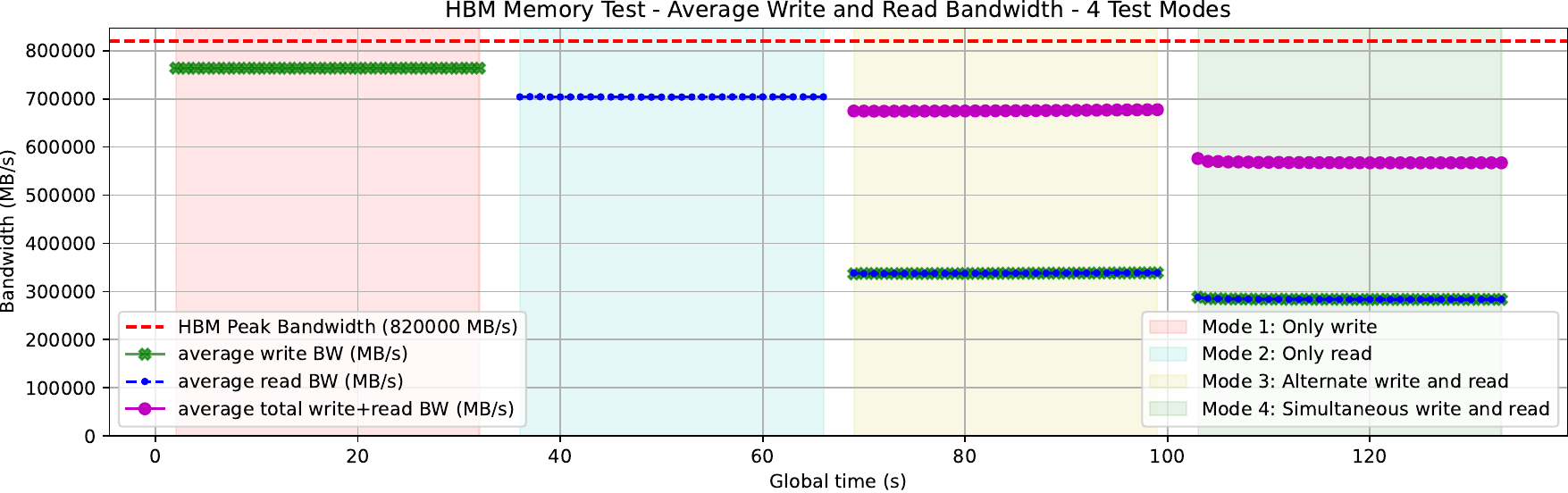}
\caption{Measured read and write performance of the High-Bandwidth Memory over 64 pseudo-channels ($2$ stacks $\times$ $8$ memory channels $\times$ $2$ pseudo channels) in four different test modes.}
\label{fig:fpga-hbm-measurements}
\end{figure}

\begin{figure}
\centering
\includegraphics[width=1\textwidth]{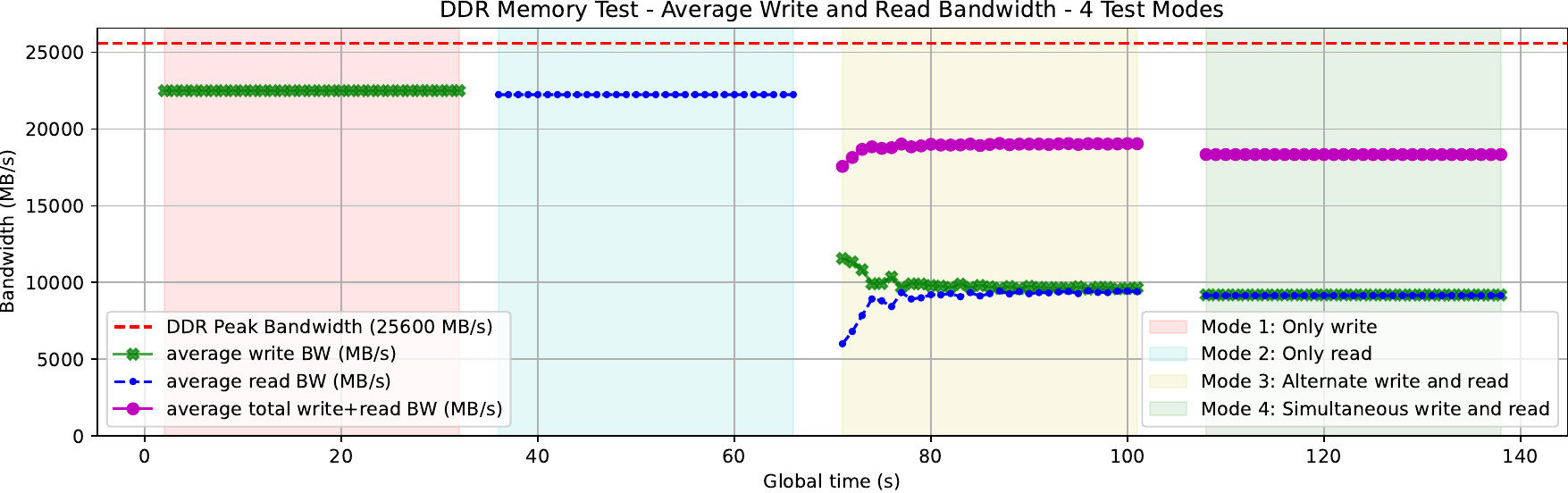}
\caption{Measured read and write performance of the DDR Memory over 2 pseudo-channels in four different test modes.}
\label{fig:fpga-ddr-measurements}
\end{figure}

The measured results are depicted in \autoref{fig:fpga-hbm-measurements} for HBM and in \autoref{fig:fpga-ddr-measurements} for DDR4. Each test mode runs for \SI{30}{\second}. 

The HBM results show that, depending on the test mode (1, 2, 3 or 4), an average bandwidth of (\SI{93.1}{\percent}, \SI{85,9}{\percent}, \SI{82,6}{\percent} and \SI{69,2}{\percent}) of the theoretical peak bandwidth can be reached. For the isolated write and read test modes (1 and 2),  read performance is slightly better; however, the test modes that combine read and write (3 and 4) achieve similar performance for the individual operations. 

The DDR memory results show a similar trend, but with important differences. On average, the test modes can reach a bandwidth of (\SI{87,9}{\percent}, \SI{86,9}{\percent}, \SI{74,4}{\percent} and \SI{71,6}{\percent}) of the theoretical peak. In contrast to HBM, the isolated write and read tests (modes 1 and 2) reach a very similar performance. The alternating write and read (test mode 3) reaches a steady state after one-third of the execution time. The simultaneous write and read (mode 4) behaves similarly to the HBM case.

\subsection{Toolflows and Board Support Package}
\label{sec:fpga-infrastructure:bsp}

For FPGA design development, the focus is on supporting high-level synthesis (HLS) toolflows. In contrast to traditional hardware description languages (HDLs), both the host software and FPGA accelerator kernels are written in high-level languages such as C++. This approach, together with the availability of efficient compilers and libraries, lowers the barrier to FPGA adoption and increases developer productivity. 

On \NOCTUATWO, the Alveo U280 cards are programmed with Xilinx Vitis HLS and the runtime is managed by Xilinx XRT (user space libraries and kernel drivers to handle the interaction between host and FPGA). The BittWare 520N cards are programmed with oneAPI and use the SYCL/DPC++ runtime. For the V80 cards in \OTUS, XRT is not available; instead, the cards are programmed and managed with SLASH \citep{SLASH-VRT}. SLASH comprises two main components: 1) a base board support package (BSP) that layers on top of AVED \citep{AVED}, and 2) VRT that provides the runtime and kernel drivers. SLASH/VRT is deployed on all nodes and the usage is documented in the getting started guide\footnote{\url{https://upb-pc2.atlassian.net/wiki/spaces/PC2DOK/pages/607125508/Otus+FPGA+Pilot+Phase}}.

\paragraph{Board Support Package Sizes}
As outlined above in \autoref{sec:fpga-infrastructure:hw-details}, the device foundation of the V80 card is fundamentally different from a traditional FPGA fabric and integrates extensive hardened I/O and protocol blocks. To estimate how this affects the size of the static BSP, the resource usage of the BSPs employed in \NOCTUATWO is compared with that of the new SLASH/VRT BSP. The overview is presented in \autoref{tab:fpga-bsp-size}. The following notes should be taken into account when interpreting these numbers:

\begin{itemize}
  \item[*] The VRT/AVED card management layer currently does not support partial reconfiguration. The reported figures therefore correspond to the exact resources used by the BSP and will likely increase with future updated versions, where additional rectangular regions around the BSP resources will be reserved to enable partial reconfiguration.
  \item[**] These numbers are obtained from \textit{platforminfo} and represent all resources in regions reserved by the BSP. When using HBM, additional resources are required for the HBM subsystem and are counted as part of the user kernel region.
  \item[***] These Numbers are collected from \textit{board\_spec.xml} and designate all resources in regions blocked by the BSP.
\end{itemize}

\begin{table}
\centering
\begin{tabular}{|rrrr|}
\multicolumn{4}{l}{\textbf{BSP FPGA Resources Utilization}} \\
\hline
            & Xilinx Alveo V80*  & Xilinx Alveo U280**                 & BittWare 520N***                       \\
\hline
toolflow            & VRT/AVED  & XRT                 & oneAPI                       \\
version           & gen5x8\_24.1  & gen3x16\_xdma\_1\_202211\_1                 & p520\_max\_sg280l/20.4                       \\
\hline
Logic                &  \textless \SI{2}{\percent}  & \SI{13.1}{\percent}     & \SI{25.2}{\percent}  \\
\myrowcolour
DSPs                & \textless \SI{1}{\percent} & \SI{11.9}{\percent}    & \SI{18.2}{\percent}  \\
BRAM           & \textless \SI{2}{\percent}  & \SI{8.0}{\percent} & \SI{23.6}{\percent}  \\
\hline

\end{tabular}
\caption{Comparison of the FPGA BSP sizes.}
\label{tab:fpga-bsp-size}
\end{table}

Even though the current values are not directly comparable, the extensive hardened blocks are expected to have a noticeable impact, which will be quantified more precisely in future work.

\subsection{Cluster-Level Integration}
This section provides an overview of FPGA integration at the system level. The general setup builds on the integration established in \NOCTUATWO. Please refer to the original work \citep{bauer2024noctua} for more details. In the current pilot phase, not all concepts have been migrated to \OTUS. Two main use cases are important for the system-level integration of FPGAs in HPC and need to be supported:

\paragraph{1. Design Development}
Users require a consistent system environment to develop, compile/synthesize, and emulate their designs on any node of the cluster. Pre-installed programming toolchains are provided and centrally maintained and updated. In order to use the tools, a dedicated Lmod gateway \textit{fpga} is available to all users that also handles dependencies, driver, and utility loadings. An example of the overall structure can be seen in \autoref{list:fpga-gateway}:

\lstset{
basicstyle=\footnotesize\ttfamily,
language=Bash,
captionpos=b,
numbers=none,
framexleftmargin=0mm,
xleftmargin=0em,
morecomment=[l]{\#} 
}
{
\begin{lstlisting}[
float=ht,
escapeinside={/*@}{@*/},
frame=tb,
commentstyle=\color{listComment},
keywords={if,then,while,else,end,or,and,def,return,srun},
keywordstyle=\color{listKeyword},
label=list:fpga-gateway, 
caption=FPGA gateway module to provide software tools and utilities.
]
$ module load fpga              # Load FPGA gateway module.
$ module available              # Show available modules.

# List of available development toolflows. (D) = Default version to load.
xilinx/vitis/24.1   xilinx/vitis/24.2   xilinx/vitis/25.1  (D) 
[...]

# List of FPGA utilities, for example to use direct FPGA-to-FPGA networking.
intel/testFPGAlinks
intel/channel_emulation_patch
changeFPGAlinks
\end{lstlisting}
}

For the FPGA synthesis, which involves translating HLS or HDL code into an actual FPGA configuration, users can set a higher priority for their jobs using the \textit{\#SBATCH -q fpgasynthesis} quality-of-service feature of Slurm. FPGA synthesis often involves testing multiple options through design space exploration to achieve a balanced final design. This process is very memory-intensive and can take from hours to days to complete. The approach of giving these jobs  a higher priority has proven effective in the previous systems and will continue in the same way. All nodes of \OTUS provide the necessary software to run synthesis jobs.

\paragraph{2. Hardware Execution}
Users should only need to allocate FPGA nodes for actual hardware execution. As described in \autoref{sec:fpga-infrastructure:bsp}, the FPGA cards are configured with a specific BSP version. On \NOCTUATWO, functionality has been integrated into the Slurm workload manager to request FPGA nodes with the appropriate configuration. An example job script illustrating this mechanism is shown in \autoref{list:fpga-base-bitstream}.

\lstset{
basicstyle=\footnotesize\ttfamily,
language=Bash,
captionpos=b,
numbers=none,
framexleftmargin=0mm,
xleftmargin=0em,
morecomment=[l]{\#} 
}
{
\begin{lstlisting}[
float=ht,
escapeinside={/*@}{@*/},
frame=tb,
commentstyle=\color{listComment},
keywords={if,then,while,else,end,or,and,def,return,srun},
keywordstyle=\color{listKeyword},
label=list:fpga-base-bitstream, 
caption=Allocating FPGA Node with specific base bitstream. Example from \NOCTUATWO.
]
/*@\colorbox{blue!7}{\tiny{By requesting a node in the FPGA partition with the constraint BittWare\_520n\_20.4.0\_max, }}@*/
/*@\colorbox{blue!7}{\tiny{the base bitstream and driver for the BittWare card in version 20.4.0 will be provided for this job.}}@*/
srun --partition=fpga --constraint=BittWare_520n_20.4.0_max ./fpga_appl
\end{lstlisting}
}

Users must provide a \textit{-{}-constraint} argument along with the required base bitstream and version. Both must match the version that the design was synthesized with. The workload manager then allocates a free node that matches the requested configuration, or configures a free node with the requested configuration. As this functionality is not currently required in the pilot phase, it will be migrated from \NOCTUATWO to \OTUS in the future.

\section{Software Stack, Services and System Management}
\label{otus-software-and-system-management}
\OTUS uses Rocky Linux as the operating system on all CPU, GPU and FPGA nodes, which is binary-compatible with Red Hat Enterprise Linux (RHEL). Login and admin nodes run with RHEL. The job scheduling software is Slurm~\citep{yoo2003slurm}, and pre-installed software is provided with Lmod~\citep{mclay2011lmod}. \autoref{list:gateway-modules} gives an overview of the available groups of modules (gateways). Further technical details can be found in the \NOCTUATWO paper~\citep{bauer2024noctua}. 

\lstset{
basicstyle=\footnotesize\ttfamily,
language=Bash,
captionpos=b,
numbers=none,
framexleftmargin=0mm,
xleftmargin=0em,
morecomment=[l]{\#} 
}
{
\begin{lstlisting}[
float=ht,
escapeinside={/*@}{@*/},
frame=tb,
commentstyle=\color{listComment},
keywords={if,then,while,else,end,or,def,return,srun},
keywordstyle=\color{listKeyword},
label=list:gateway-modules, 
caption={
List of available gateway modules with pre-installed software (D: default module; G:  gateway module; L: module is loaded *: module built for host, native FPGA and offload to FPGA})]
   /*@\colorbox{blue!7}{-{}-{}-{}-{}-{}-{}-{}-{}-{}-{}-{}-{}-{}-{}-{}- Gateway- and basic modules -{}-{}-{}-{}-{}-{}-{}-{}-{}-{}-{}-{}-{}-{}-}@*/
   DefaultModules (S,L)    fpga   (*,G)    slurm/24.11.5-1
   all            (G)      lang   (G)      slurm/24.11.6-1 (L,D)
   bio            (G)      lib    (G)      system          (G)
   cae            (G)      math   (G)      toolchain       (G)
   chem           (G)      mpi    (G)      tools           (G)
   compiler       (G)      numlib (G)      vis             (G)
   data           (G)      pc2env (L)
   devel          (G)      perf   (G)
\end{lstlisting}
}

\subsection{Job Monitoring}
\label{sec:jobmonitoring}
\begin{figure}[h!]
\centering
\includegraphics[width=1\textwidth]{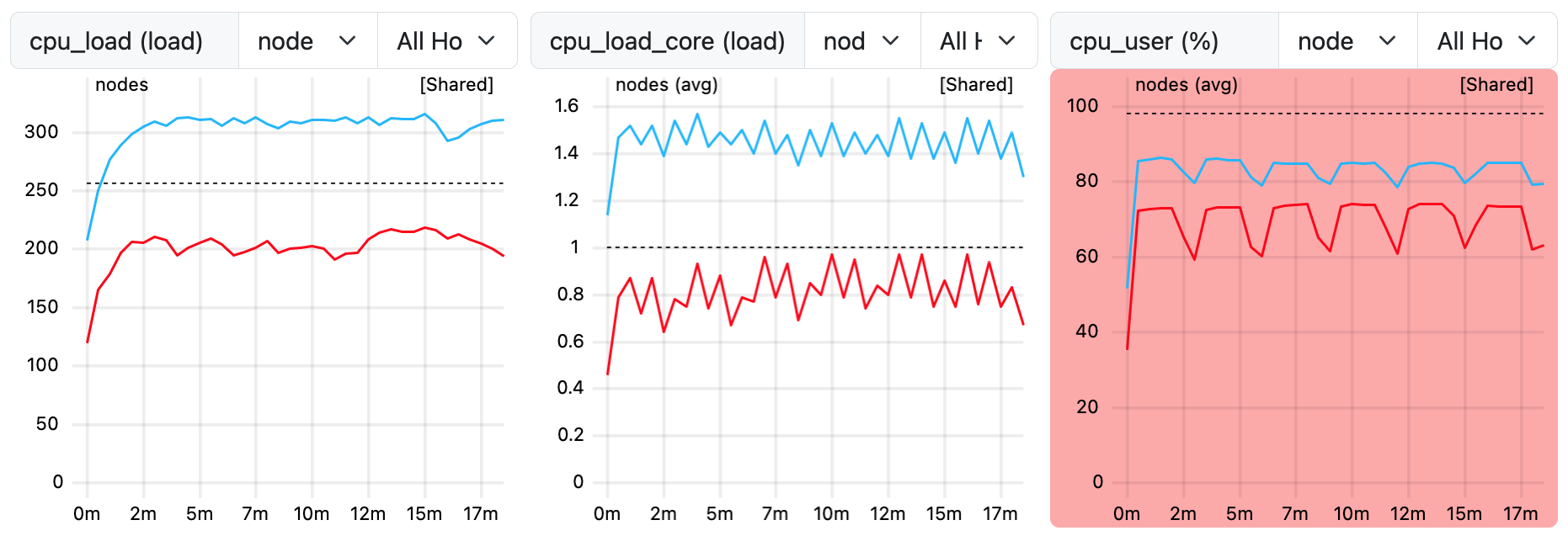}
\caption{
ClusterCockpit job monitoring example plots showing CPU load (running/runnable threads on the complete system) (left), per-core load (middle), and CPU user utilization (right) over time. The higher blue line compared to the lower red line indicates a persistent load imbalance between two compute nodes. Per-core load values exceeding $1.0$ (middle plot) reveal CPU core over-subscription.
}
\label{fig:jobmonitoring}
\end{figure}

A job-oriented monitoring framework called \mbox{ClusterCockpit}~\citep{eitzinger2019clustercockpit} is provided to give users the opportunity to analyze performance issues of their jobs. 
The ClusterCockpit continuously records various metrics on the compute nodes and links them to the key job data. Users can then view the relevant metrics for their job in a web interface. They can sort the list by different aspects (e.g. CPU load, main memory usage, memory bandwidth) by clicking on the appropriate buttons, and display or hide metrics or apply filters to the list. As depicted in \autoref{fig:jobmonitoring}, the metrics are plotted over time, enabling quick recognition of performance behavior. GPU jobs are also included in the job monitoring and users can inspect GPU metrics like compute utilization, memory utilization, and many more.
This service enables support staff and users to examine application behavior, even long after the computer job has finished.
Efficiency problems can be identified and solutions can be evaluated.
Without job monitoring, program debugging and analysis would be more difficult.

Compared to Zen 3 CPUs in \NOCTUATWO, Zen 5 CPU architecture in \OTUS supports more detailed monitoring events.
Memory bandwidth can now be differentiated between read and write memory operations.
The floating-point operations rate is also available separately for double and single precision.
The parallel file system metrics have been adapted to the file system used in \OTUS.
These additional metrics provide improved insight into the behavior of user calculations and performance characteristics.
The remaining metrics are familiar from the \NOCTUATWO job monitoring service.

\subsection{System Management}

\OTUS uses three admin nodes to manage the cluster and provide services such as DHCP, DNS, NTP, NFS, Slurm, fabric management, monitoring and log aggregation. Pacemaker \citep{PCM} and Corosync \citep{CORS} are used for high availability and load balancing between the three admin nodes, meaning that each head node runs a subset of services. Storage and NFS for the cluster and the services are provided by a separate Lenovo ThinkSystem DM3000H Hybrid Storage Appliance. \\

The admin nodes are Lenovo ThinkSystem SR635 V3 with AMD EPYC Genoa 9454 processors (48 cores/96 threads), \SI{512}{\giga\byte} of memory, dual-port 100 GbE for internal networking, dual-port 10/25 GbE for external networking, dedicated GbE BMC-port, dual-port NDR200-IB, \SI{1}{\tera\byte} SATA SSD. \newline

The entire cluster is configured using Lenovo Confluent \citep{CFL} which is a Python3-based cluster manager. Confluent handles the bootstrapping and configuration of the nodes. It is the modern successor of xCAT \citep{xCAT}. \\

The cluster nodes boot predefined images. These images are created and managed with Confluent. The workflow is as follows: First, create a bare image containing only the operating system. Then, mount the bare image and roll out the necessary changes. The image is then packed, compressed and unmounted. Finally, a node can be selected to boot from the newly created image. \newline

\subsection{User Portal}
\label{sec:userportal}

The level of experience and knowledge among HPC users can vary widely. \PCTWO provides a range of HPC-related services, along with the corresponding documentation. To enhance the value of \PCTWO resources for the users, a central hub called the \textit{User Portal} is offered. The portal is reachable via a web browser and presents useful information, functionality and links for users in all stages of their project life cycle in one single location. The following paragraphs will explain the most important features of the portal.

\begin{figure}[h!]
\centering
\includegraphics[width=1\textwidth]{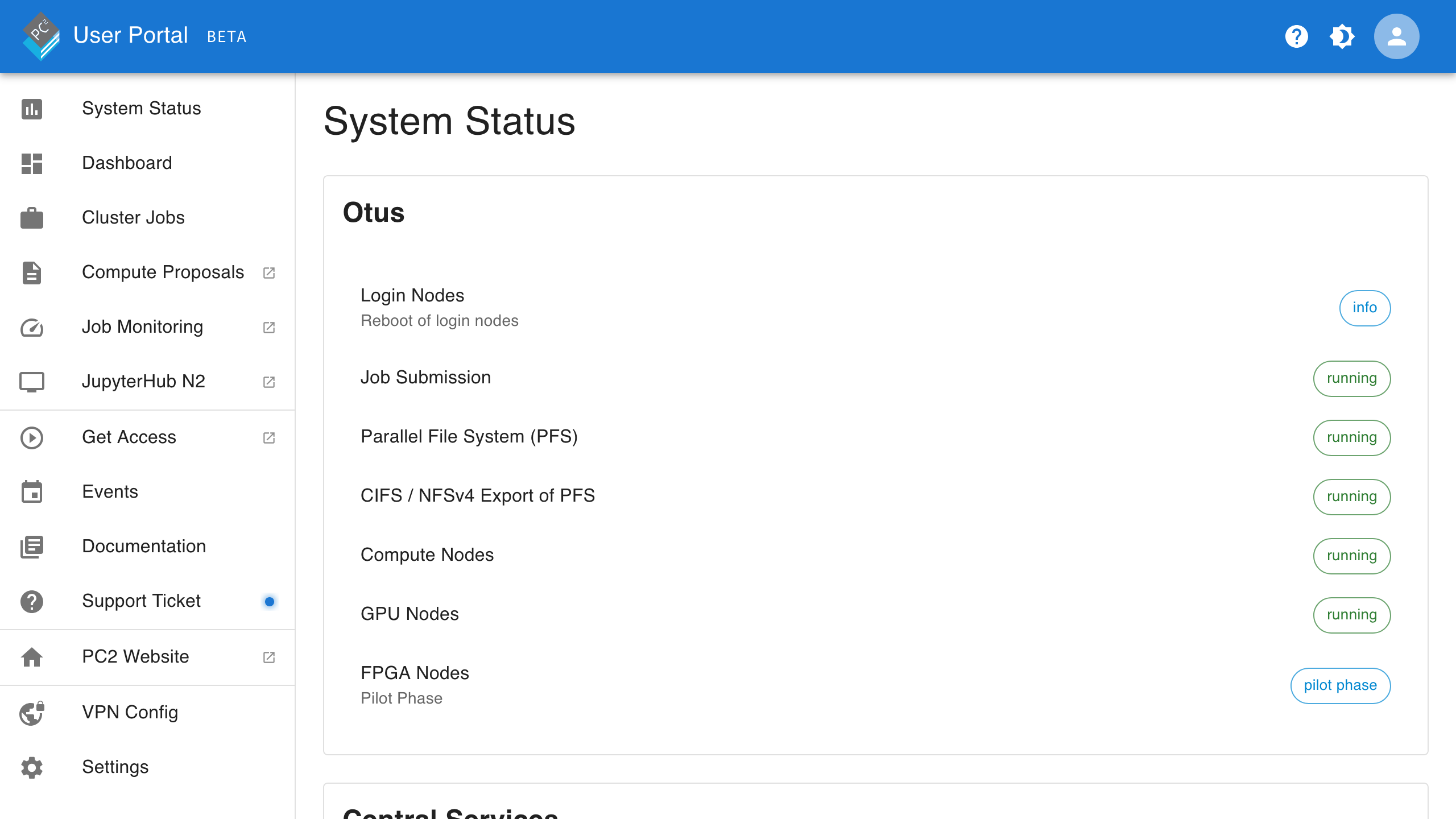}
\caption{System status overview from the User Portal web app.}
\label{fig:portal-system-status}
\end{figure}

\paragraph{System Status}
The system status page displays key information about all operational cluster systems and central services, including planned maintenance downtime, reduced availability and minor service issues. Each compute cluster is divided into subcategories such as node types (login, cpu-only, gpu, fpga, etc.), parallel file system, networking, and other services. The current status of each category is visible at a glance and previous messages can be accessed via a history of events. There is also an option for \PCTWO staff to add important alerts that are displayed on each portal subpage, not just on the status page. This status page is the default view when accessing the User Portal, see \autoref{fig:portal-system-status}.

\begin{figure}
\centering
\includegraphics[width=1\textwidth]{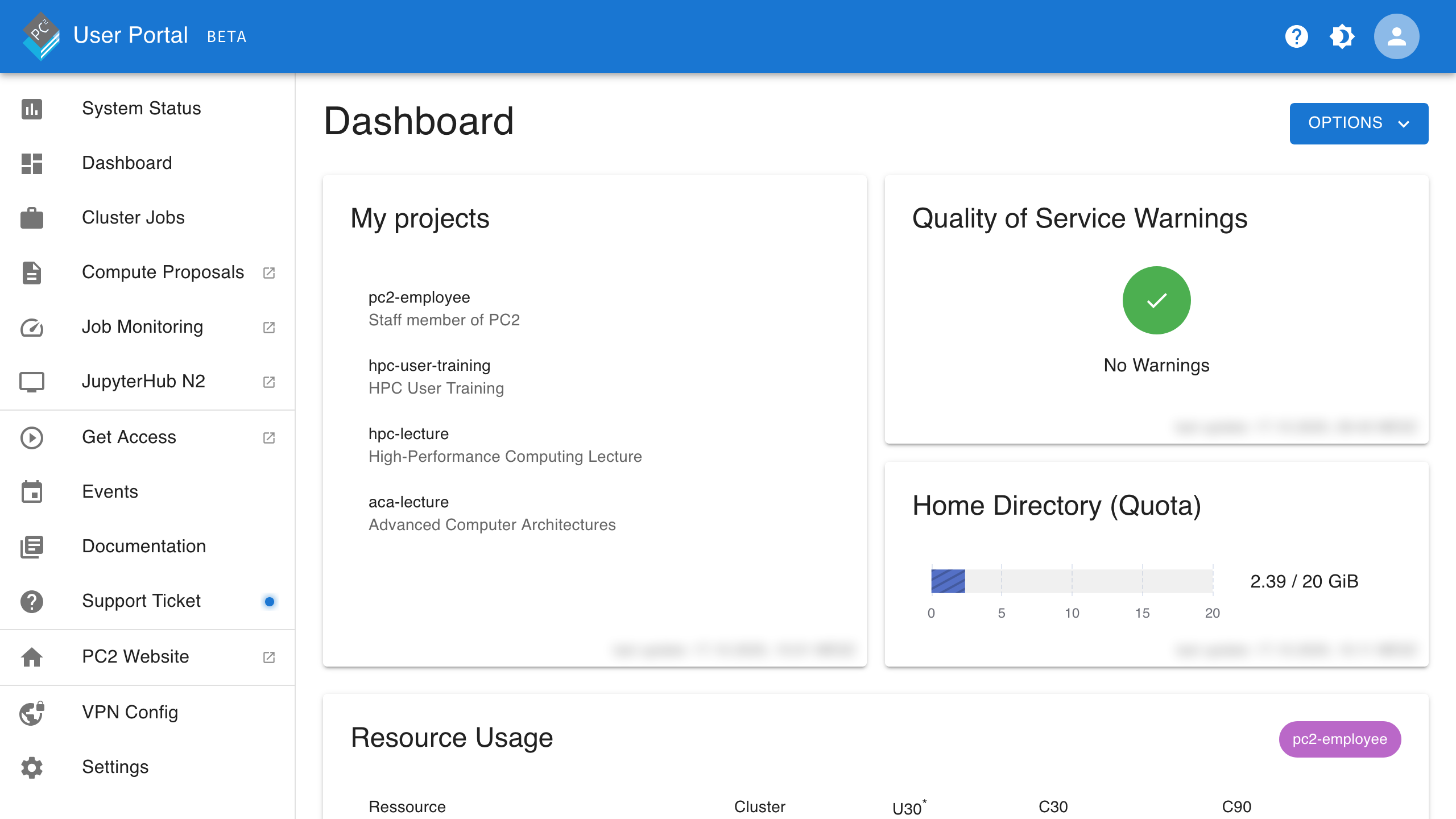}
\caption{User configurable dashboard showing various types of information like project membership, file system quotas, resource usage, resource warnings, etc.}
\label{fig:portal-dashboard}
\end{figure}

\paragraph{Dashboard}
The dashboard page (see \autoref{fig:portal-dashboard}) can display the user’s project information. This includes project runtimes (total, elapsed and remaining), resource usage, and warnings for all accessible cluster systems as well as file system quotas for the parallel file systems and the user’s home directory. After logging in, users can access and configure the dashboard, e.g. by adding	a system status information widget.

\paragraph{Jobs}
The user’s pending, running and recently finished cluster jobs from the Slurm workload scheduler can be viewed on a dedicated subpage of the portal. Jobs from projects where the user is a coordinator can also be displayed to provide a comprehensive view of the supervised projects. The table containing these jobs can be searched, filtered and sorted. Clicking on the respective entry reveals additional metadata for each job. The selected job can then be opened directly in the ClusterCockpit job monitoring service for in-depth analysis (e.g. resource usage of the requested CPUs and GPUs), see \autoref{sec:jobmonitoring}. A support request can also be created directly for the selected job in case any issues arise.

\paragraph{Other Services}
Separate services such as the JARDS proposal system, JupyterHub and the ClusterCockpit job monitoring system can be easily accessed from the User Portal. Many of these services offer single sign-on, so no additional authentication is required beyond the initial portal login.

\paragraph{Documentation and Training}
Useful for all users, especially the newer ones, is the collection of links to information sources like \PCTWO webpages, \PCTWO-maintained documentation and FAQs, and other wikis and tutorials. Information that is often needed is highlighted, e.g. how to get access to the resources, how to get started on the cluster and start jobs, which software is pre-installed, etc. A dedicated page shows upcoming events like training sessions and workshops with various topics, consultation sessions and other special events.

\paragraph{Support}
For support requests, another page of the portal guides the users through how to write a good support ticket and provides a web form with partially prefilled information, which can then be sent off to \PCTWO. Responses will be sent to the email address stored in the corresponding user account.

\paragraph{Self-Service}
The user portal enables users to perform an increasing number of tasks related to cluster access themselves. To this end, users can already upload their SSH key automatically to access the cluster via the user portal settings. They can also set up the necessary VPN as a self-service. The growing range of self-service options is intended to reduce the workload of data center staff and provide users with a faster way to perform common administrative tasks.

\clearpage
\subsection{Compute Project Management System PERSEUS}
\label{sec:perseus}

Scientific HPC centers operate in a dynamic environment and must adapt their internal processes relatively frequently. For a long time, there has been no suitable open-source software available that addresses this problem and enables these centers to flexibly configure their compute project processes. \\

\begin{figure}[h!]
\centering
\includegraphics[width=1\textwidth]{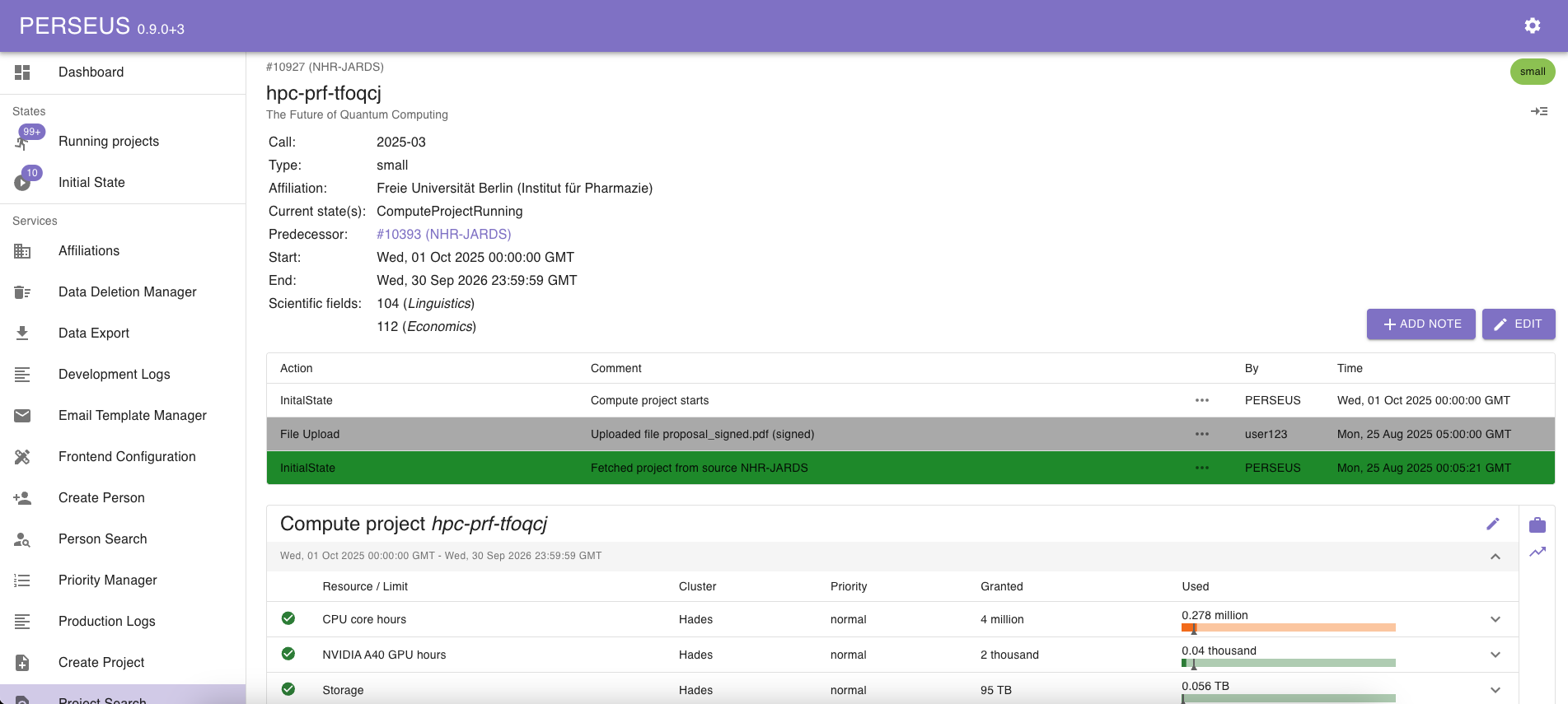}
\caption{Screenshot from the detailed view for a project in PERSEUS. The demo mode was used to create mock data for the database.}
\label{fig:perseus-screenshot}
\end{figure}

\PCTWO provides a free and open-source solution with PERSEUS\footnote{\url{https://pc2.uni-paderborn.de/perseus}} that supports data and workflow management for compute projects, as well as center-wide automation while being fully customizable. The PERSEUS project is licensed under the MIT License\footnote{\url{https://opensource.org/license/mit}}. \autoref{fig:perseus-screenshot} gives an overview of the management functionalities in the left menu and a detailed view of a compute project in the middle.

\begin{figure}
\centering
\includegraphics[width=1\textwidth]{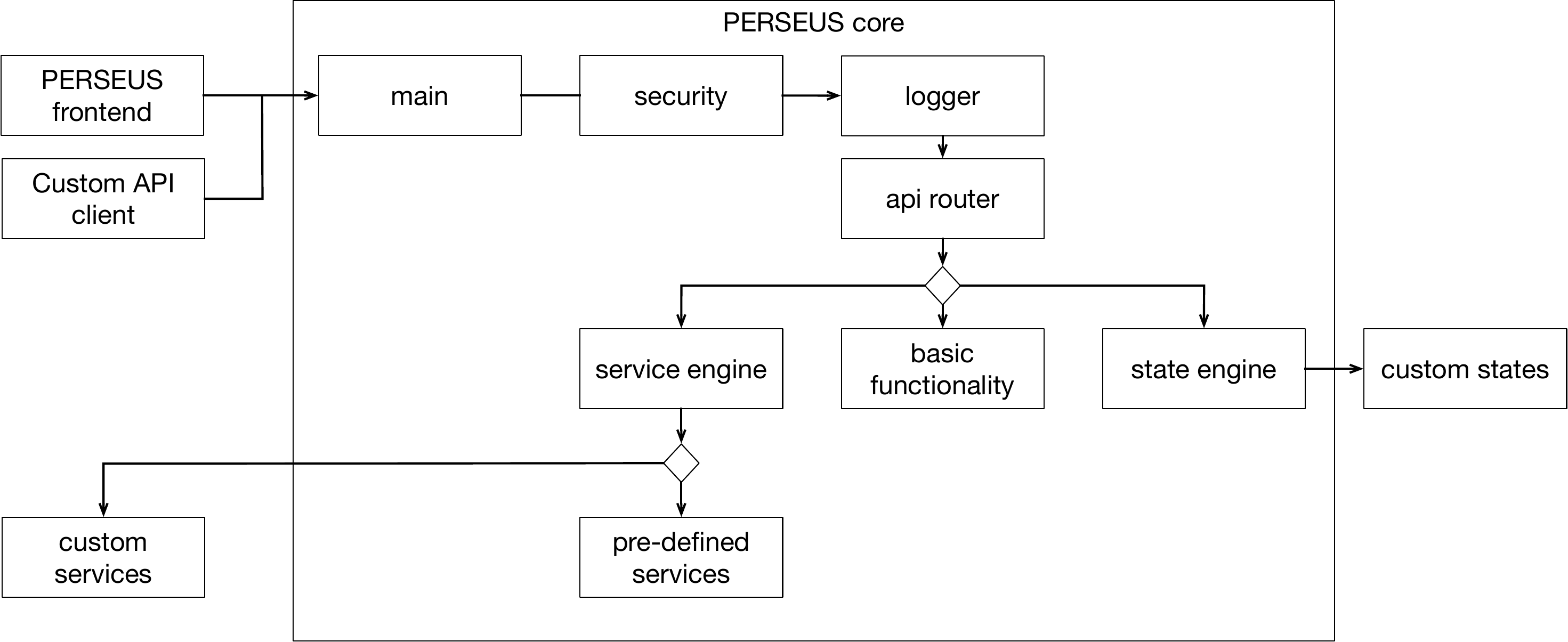}
\caption{Request structure for PERSEUS core. User-defined services and states are located outside the core component, allowing for customization without risking the updatability.}
\label{fig:perseus-request-structure}
\end{figure}

\paragraph{Architecture}
PERSEUS comprises two software components: the frontend and the core. The frontend, which was developed as a React app with TypeScript, enables HPC center employees to access all functionalities quickly and easily via a web interface. The core component, which is a FastAPI application developed with Python, contains the actual logic. Apart from basic functionality such as authentication, every call to the PERSEUS core ends up either in the state engine or the service engine (see \autoref{fig:perseus-request-structure}).

\paragraph{States}
Each state represents a specific condition in which a compute project may be found, such as undergoing formal review or being subject to export control. Some states only perform automatic processes, while others require human interaction. For this reason, the states can output configurable tasks that must be completed by HPC center staff. It is also possible to configure how certain inputs in the forms for these tasks manipulate project data or influence subsequent workflows.\\
To use these states, a state machine must be assigned to each new project. This defines the initial state, all possible state transitions and, if applicable, the final states, thereby formalizing the project's life cycle and fully automating its processing.

\paragraph{Services}
Services provide general functionality that is not assigned to a specific state. PERSEUS comes with several standard services, such as a project search or a project editor.\\
Services can provide API endpoints that can be used for automation purposes. For instance, self-services from the user portal (see \autoref{sec:userportal}) can automatically edit project data or trigger other events.\\
Services can also register cron jobs that initiate recurring processes, such as regularly pulling new project applications from application portals on a regular basis, or synchronizations of any kind with other systems.

\paragraph{Customizability}
Both states and services are added as PERSEUS plugins via a container volume. Each plugin is represented by Python files that can either be developed from scratch or downloaded from the public collection on GitHub\footnote{\url{https://github.com/pc2-perseus}}.\\
This ensures that required functionality can be added quickly and easily without compromising the updatability of PERSEUS itself. When an update is performed, the container image can simply be renewed; the plugins continue to be loaded via the volume in PERSEUS core.

\clearpage
\section{Data Center Building, Power and Cooling Infrastructure}
\label{sec:noc2:power-and-cooling}

\begin{figure}
\centering
\includegraphics[width=1\textwidth]{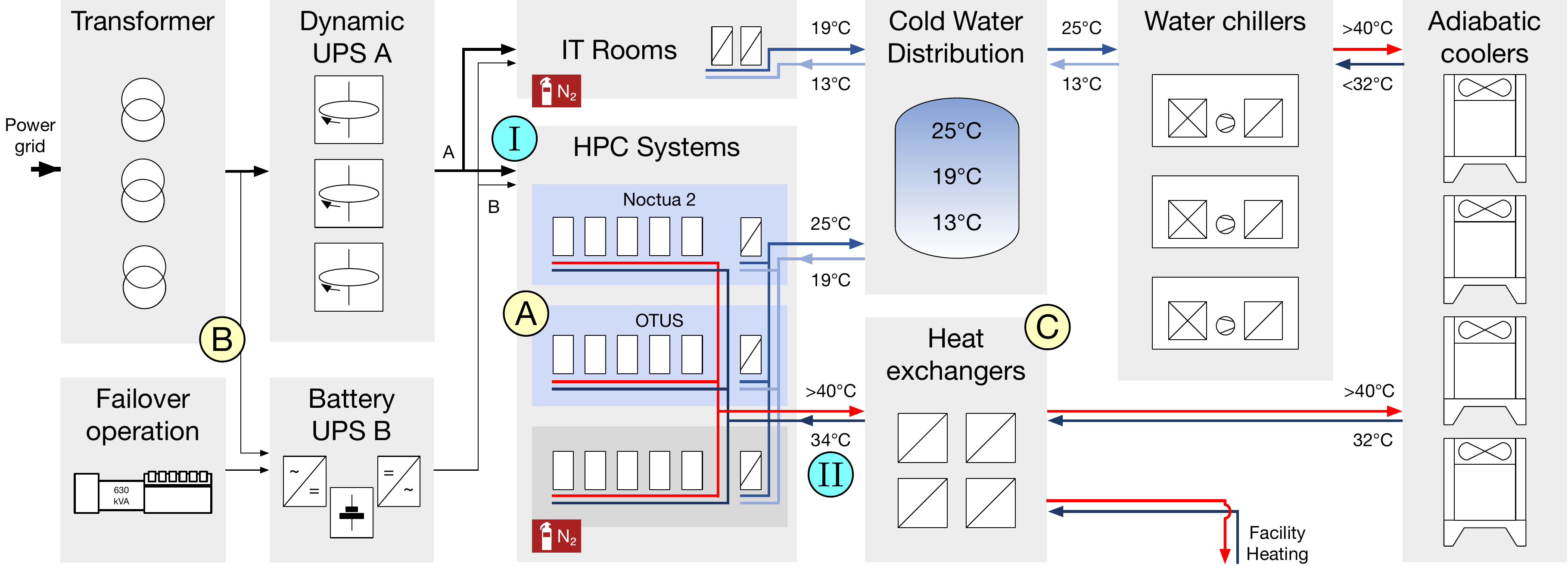}
\caption[Schematic depiction of the power and water cooling infrastructure present in order to operate \OTUS and \NOCTUATWO. \OTUS is connected to the facility power delivery at point \textit{I} and the warm water loop at point \textit{II}.]{
Schematic depiction of the power and water cooling infrastructure present in order to operate \OTUS and \NOCTUATWO. \OTUS is connected to the facility power delivery at point \circledcyan{\uproman{I}} and the warm water loop at point \circledcyan{\uproman{II}}.}
\label{fig:n2_power_and_cooling}
\end{figure}

\autoref{fig:n2_power_and_cooling} shows the overall view of the data center building operated by \PCTWO. The entire technical setup is described in detail in the \NOCTUATWO paper \citep{bauer2024noctua}. In this section, only the highlights of the overall setup and updates compared to the last described status are described. 

The data center building is designed for a highly energy-efficient operation and effective waste heat reuse. Combining cold and hot water cooling enables us to run compute units within their specifications while recovering at least \SI{85}{\percent} of the waste heat from the hot water cooling for reuse in campus heating. Power Usage Effectiveness (PUE) is a key metric for estimating the energy efficiency of a data center infrastructure. PUE is calculated by dividing the total energy by the energy used for IT equipment. The closer the value is to \SI{1.0}{}, the more energy efficient the data center is and the better its energy balance. During the evaluation in the first month of operation, a PUE value of \SI{1.122}{} was recorded for \OTUS, which is excellent by industry standards. 

Regarding updates to the infrastructure, first, the white space that can accommodate three HPC systems in separate segments has been extended by the \OTUS cluster \circled{A}. One segment is still free for another extension. Second, the expansion of the electrical and cooling capacities \circled{B} and \circled{C} has been started and is under construction. By the end of 2026 the electrical capacity will be increased from \SI{2.6}{\mega\watt} to \SI{4.6}{\mega\watt} and an option for another extension to a total of \SI{6}{\mega\watt} is still available, if required. The cooling capacities will be expanded accordingly.

\section{Conclusion}
\label{sec:conclusion}
In conclusion, this article provides an overview of the \OTUS supercomputer and offers detailed insights into its efficient operation at the Paderborn Center for Parallel Computing (\PCTWO). \OTUS is the successor to \NOCTUATWO, offering approximately twice the computing power while maintaining a similar configuration to facilitate ease of use and migration for those familiar with the previous system.

The article discusses the different node types and gives detailed performance measurements. Furthermore, it outlines the software stacks, services, and system management setup to provide insights into the system configuration. Finally, the integration of \OTUS within the data center building is described, highlighting the infrastructure that is required for reliable, efficient operation.

\section*{Acknowledgments}
The authors gratefully acknowledge the joint support by the Federal Ministry of Research, Technology and Space and the state governments participating in the National High-Performance Computing (NHR) joint funding program\footnote{\url{www.nhr-verein.de/en/our-partners}}. We also thank AMD/Xilinx for the support as part of the Heterogeneous Accelerated Compute Clusters (HACC) program\footnote{\url{https://www.amd-haccs.io}}.

\section{Appendix}
\label{sec:appendix}
This appendix provides supplementary material.

\subsection{System Setup Visuals}
\label{sec:appendix-visuals}

\begin{figure}
\centering
\includegraphics[width=1\textwidth]{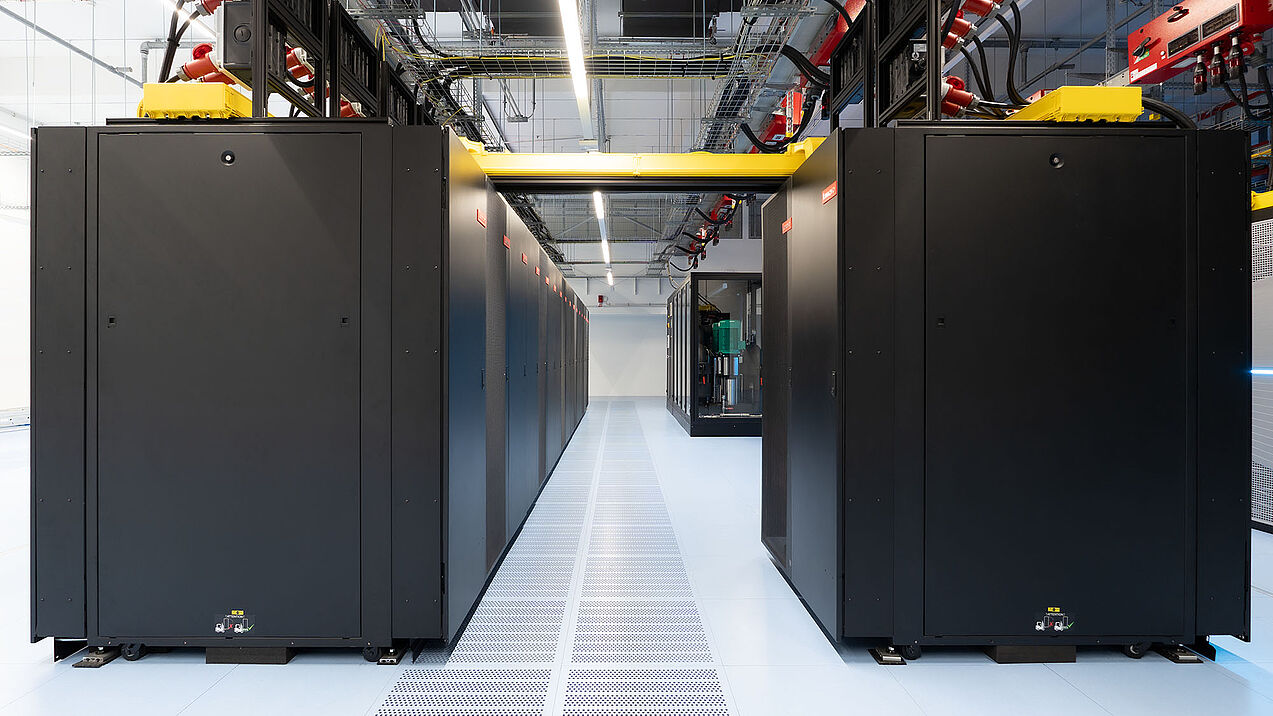}
\caption{
Side view of \OTUS supercomputer. Five pods with the additional FPGA racks on the left side. On the right side are the two additional racks in the front and the Cooling Distribution Unit (CDU) in the back.}
\label{fig:otus_side}
\end{figure}

\begin{figure}
\centering
\includegraphics[width=1\textwidth]{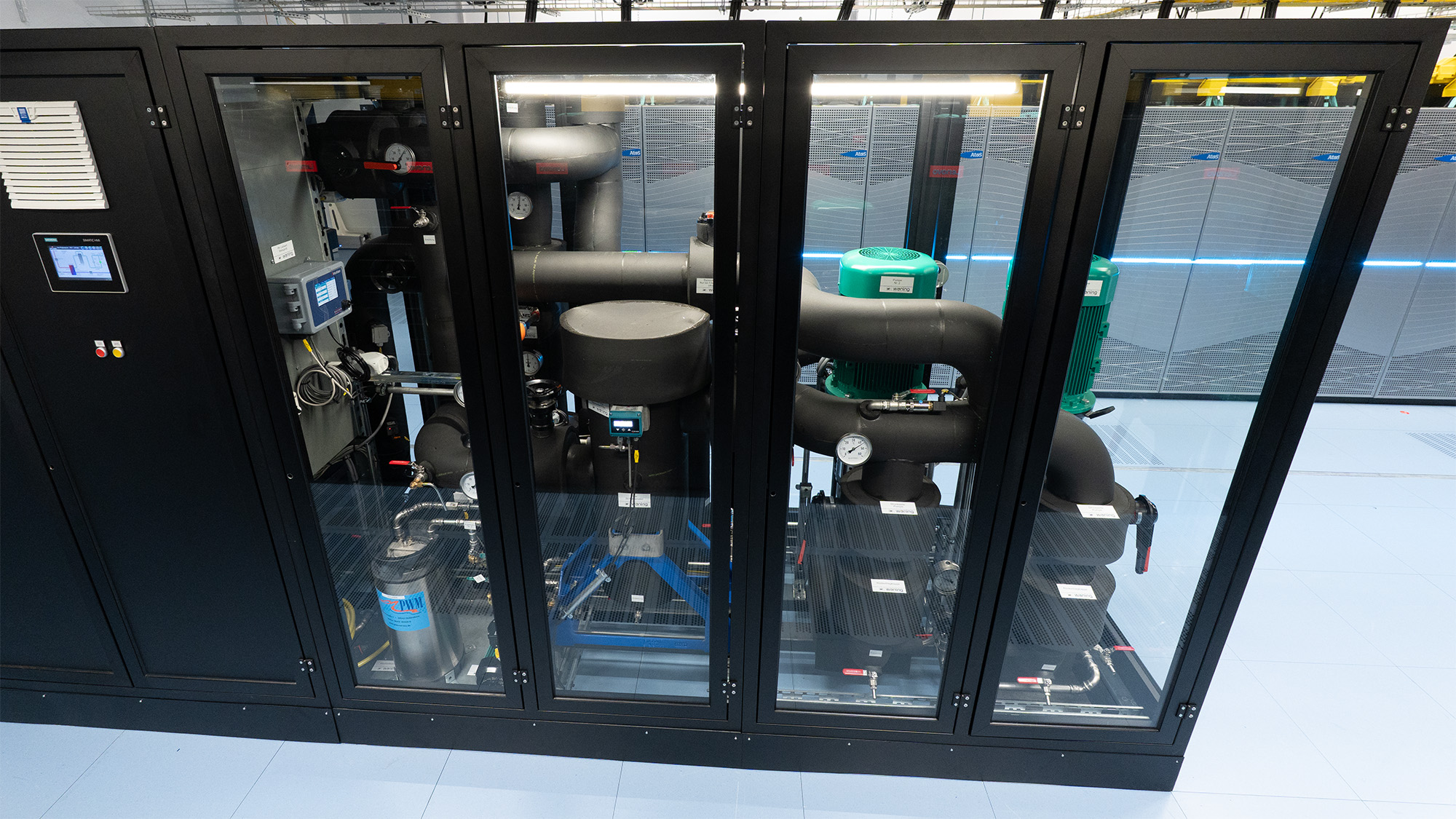}
\caption{
Central Cooling distribution unit of \OTUS supercomputer. All connecting pipes are located below the raised floor.}
\label{fig:otus_cdu}
\end{figure}

\clearpage
\bibliographystyle{plainnat}
\bibliography{biblio}

@misc{devices2020software,
  author =        {AMD},
  publisher =     {AMD},
  title =         {56665, Software Optimization Guide for AMD Family 19h
                   Processors (PUB)},
  year =          {2020},
}

@misc{AVED,
  author =        {AMD},
  howpublished =  {[Online]
                       \url{https://xilinx.github.io/AVED/latest/index.html}},
  note =          {Accessed: 2025-10-15},
  title =         {{AMD Alveo Versal Example Design (AVED)}},
  year =          {2025},
}

@misc{Alveo-DS1013,
  author =        {AMD},
  howpublished =  {[Online]
                   \url{https://docs.amd.com/r/en-US/ds1013-v80}},
  note =          {Accessed: 2025-10-15},
  title =         {{Alveo V80 Data Sheet (DS1013)}},
  year =          {2025},
}

@misc{SLASH-VRT,
  author =        {AMD},
  howpublished =  {[Online] \url{https://github.com/Xilinx/SLASH/}},
  note =          {Accessed: 2025-10-15},
  title =         {{SLASH VRT}},
  year =          {2025},
}

@misc{versal-pg313,
  author =        {AMD},
  howpublished =  {[Online]
  \url{https://docs.amd.com/r/en-US/pg313-network-on-chip/HBM-Topology}},
  note =          {Accessed: 2025-10-15},
  title =         {{PG313 (v1.1)}},
  year =          {2025},
}

@misc{xbtest,
  author =        {AMD},
  howpublished =  {[Online]
  \url{https://xilinx.github.io/AVED/latest/xbtest/user-guide/source/index.html}},
  note =          {Accessed: 2025-10-15},
  title =         {{Xbtest Userguide}},
  year =          {2025},
}

@article{bauer2024noctua,
  author =        {Bauer, Carsten and Kenter, Tobias and Lass, Michael and
                   Mazur, Lukas and Meyer, Marius and Nitsche, Holger and
                   Riebler, Heinrich and Schade, Robert and
                   Schwarz, Michael and Winnwa, Nils and others},
  journal =       {Journal of large-scale research facilities JLSRF},
  number =        {1},
  title =         {Noctua 2 supercomputer},
  volume =        {9},
  year =          {2024},
}

@misc{cnclatency,
  author =        {Chips and Cheese},
  howpublished =  {[Online]
  \url{https://github.com/ChipsandCheese/MemoryLatencyTest}},
  note =          {Accessed: 2025-10-15},
  title =         {{Chips and Cheese Memory Latency Test}},
  year =          {2021},
}

@misc{PCM,
  author =        {ClusterLabs},
  howpublished =  {[Online] \url{https://clusterlabs.org/pacemaker/}},
  note =          {Accessed: 2025-10-15},
  title =         {{Pacemaker}},
  year =          {2023},
}

@misc{amd2025hotchips,
  author =        {Brad Cohen and Mahesh Subramony and Mike Clark},
  institution =   {AMD},
  publisher =     {Hot Chips},
  title =         {Next Generation "Zen 5" Core},
  year =          {2024},
  url =           {https://hc2024.hotchips.org/assets/program/conference/day2/
                  24_HC2024.AMD.Cohen.Subramony.final.pdf},
}

@misc{CORS,
  author =        {{Corosync Authors}},
  howpublished =  {[Online] \url{http://corosync.github.io/corosync/}},
  note =          {Accessed: 2025-10-15},
  title =         {{Corosync}},
  year =          {2023},
}

@article{dongarra2013toward,
  author =        {Dongarra, Jack and Heroux, Michael A},
  journal =       {Sandia Report, SAND2013-4744},
  pages =         {150},
  publisher =     {Citeseer},
  title =         {Toward a new metric for ranking high performance
                   computing systems},
  volume =        {312},
  year =          {2013},
}

@inproceedings{eitzinger2019clustercockpit,
  author =        {Eitzinger, Jan and Gruber, Thomas and Afzal, Ayesha and
                   Zeiser, Thomas and Wellein, Gerhard},
  booktitle =     {{2019 IEEE International Conference on Cluster
                   Computing (CLUSTER)}},
  organization =  {IEEE},
  pages =         {1--7},
  title =         {ClusterCockpit—A web application for job-specific
                   performance monitoring},
  year =          {2019},
}

@misc{GA,
  author =        {GA},
  howpublished =  {[Online] \url{https://gauss-allianz.de}},
  note =          {Accessed: 2025-10-15},
  title =         {{Gauß-Allianz}},
  year =          {2025},
}

@article{giannozzi2009quantum,
  author =        {Giannozzi, Paolo and Baroni, Stefano and
                   Bonini, Nicola and Calandra, Matteo and Car, Roberto and
                   Cavazzoni, Carlo and Ceresoli, Davide and
                   Chiarotti, Guido L and Cococcioni, Matteo and
                   Dabo, Ismaila and others},
  journal =       {Journal of physics: Condensed matter},
  number =        {39},
  pages =         {395502},
  publisher =     {IOP Publishing},
  title =         {QUANTUM ESPRESSO: a modular and open-source software
                   project for quantum simulations of materials},
  volume =        {21},
  year =          {2009},
}

@misc{HPCNRW,
  author =        {HPC.NRW},
  howpublished =  {[Online] \url{https://hpc.dh.nrw}},
  note =          {Accessed: 2025-10-15},
  title =         {{HPC.NRW}},
  year =          {2025},
}

@misc{PC2Access,
  author =        {PC2},
  howpublished =  {[Online]
                   \url{https://pc2.uni-paderborn.de/system-access}},
  note =          {Accessed: 2025-10-15},
  title =         {{PC2 System Access Page}},
  year =          {2025},
}

@misc{gpfs,
  author =        {IBM},
  howpublished =  {[Online]
                       \url{https://www.ibm.com/de-de/products/storage-scale}},
  note =          {Accessed: 2025-10-15},
  title =         {{IBM Storage Scale}},
  year =          {2025},
}

@misc{io500,
  author =        {{IO500 Foundation}},
  howpublished =  {[Online] \url{https://io500.org/}},
  note =          {Accessed: 2025-10-15},
  title =         {{IO500}},
  year =          {2023},
}

@misc{ibmilm,
  author =        {IBM},
  howpublished =  {[Online]
  \url{https://www.ibm.com/support/pages/ibm%C2%AE-storage-scale%E2%84%A2-information-lifecycle-management-policies}},
  note =          {Accessed: 2025-10-28},
  title =         {{IBM Information Lifecycle Management Policies}},
  year =          {2025},
}

@article{kuhne2020cp2k,
  author =        {K{\"u}hne, Thomas D and Iannuzzi, Marcella and
                   Del Ben, Mauro and Rybkin, Vladimir V and
                   Seewald, Patrick and Stein, Frederick and
                   Laino, Teodoro and Khaliullin, Rustam Z and
                   Sch{\"u}tt, Ole and Schiffmann, Florian and others},
  journal =       {The Journal of Chemical Physics},
  number =        {19},
  publisher =     {AIP Publishing},
  title =         {CP2K: An electronic structure and molecular dynamics
                   software package-Quickstep: Efficient and accurate
                   electronic structure calculations},
  volume =        {152},
  year =          {2020},
}

@misc{CFL,
  author =        {Lenovo},
  howpublished =  {[Online] \url{https://github.com/lenovo/confluent}},
  note =          {Accessed: 2025-10-15},
  title =         {{Lenovo Confluent}},
  year =          {2025},
}

@article{mccalpin1995memory,
  author =        {McCalpin, John D and others},
  journal =       {IEEE computer society technical committee on computer
                   architecture (TCCA) newsletter},
  number =        {19-25},
  title =         {Memory bandwidth and machine balance in current high
                   performance computers},
  volume =        {2},
  year =          {1995},
}

@inproceedings{mclay2011lmod,
  address =       {New York, NY, USA},
  author =        {McLay, Robert and Schulz, Karl W. and
                   Barth, William L. and Minyard, Tommy},
  booktitle =     {{State of the Practice Reports}},
  publisher =     {Association for Computing Machinery},
  series =        {SC '11},
  title =         {Best Practices for the Deployment and Management of
                   Production HPC Clusters},
  year =          {2011},
  doi =           {10.1145/2063348.2063360},
  isbn =          {9781450311397},
  url =           {https://doi.org/10.1145/2063348.2063360},
}

@article{hpcc_fpga_in_depth,
  author =        {Marius Meyer and Tobias Kenter and Christian Plessl},
  journal =       {Journal of Parallel and Distributed Computing},
  pages =         {79-89},
  title =         {In-depth FPGA accelerator performance evaluation with
                   single node benchmarks from the HPC challenge
                   benchmark suite for Intel and Xilinx FPGAs using
                   OpenCL},
  volume =        {160},
  year =          {2022},
  doi =           {https://doi.org/10.1016/j.jpdc.2021.10.007},
  issn =          {0743-7315},
  url =           {https://www.sciencedirect.com/science/article/pii/
                  S0743731521002057},
}

@misc{NHR,
  author =        {NHR},
  howpublished =  {[Online] \url{https://www.nhr-verein.de}},
  note =          {Accessed: 2025-10-15},
  title =         {{Nationales Hochleistungsrechnen (NHR)}},
  year =          {2025},
}

@misc{ueabs,
  author =          {{Partnership For Advanced Computing in Europe (PRACE)}},
  howpublished =  {[Online]
                   \url{https://repository.prace-ri.eu/git/UEABS/ueabs}},
  note =          {Accessed: 2025-10-24},
  title =         {{Unified European Applications Benchmark Suite
                   (UEABS)}},
  year =          {2018},
}

@misc{PC2,
  author =        {PC2},
  howpublished =  {[Online] \url{https://pc2.uni-paderborn.de}},
  note =          {Accessed: 2025-10-15},
  title =         {{Paderborn Center for Parallel Computing (PC2)}},
  year =          {2025},
}

@article{petitet2004hpl,
  author =        {Petitet, Antoine},
  journal =       {http://www.netlib.org/benchmark/hpl/},
  title =         {HPL- a portable implementation of the
                   high-performance Linpack benchmark for
                   distributed-memory computers},
  year =          {2004},
}

@inproceedings{treibig2010likwid,
  author =        {Treibig, Jan and Hager, Georg and Wellein, Gerhard},
  booktitle =     {{2010 39th International Conference on Parallel
                   Processing Workshops}},
  number =        {},
  pages =         {207-216},
  title =         {LIKWID: A Lightweight Performance-Oriented Tool Suite
                   for x86 Multicore Environments},
  volume =        {},
  year =          {2010},
  doi =           {10.1109/ICPPW.2010.38},
}

@misc{xCAT,
  author =        {{xCAT Authors}},
  howpublished =  {[Online] \url{https://github.com/xcat2/xcat-core}},
  note =          {Accessed: 2025-10-15},
  title =         {{xCAT}},
  year =          {2024},
}

@inproceedings{yoo2003slurm,
  address =       {Berlin, Heidelberg},
  author =        {Yoo, Andy B. and Jette, Morris A. and Grondona, Mark},
  booktitle =     {{Job Scheduling Strategies for Parallel Processing}},
  editor =        {Feitelson, Dror and Rudolph, Larry and
                   Schwiegelshohn, Uwe},
  pages =         {44--60},
  publisher =     {Springer Berlin Heidelberg},
  title =         {SLURM: Simple Linux Utility for Resource Management},
  year =          {2003},
  isbn =          {978-3-540-39727-4},
}

\end{document}